\documentclass[11pt,a4paper]{article}
	
	\usepackage{mathpazo}

	\usepackage{amsmath,amssymb,amsthm,amscd}
	\usepackage{mathtools}
	\usepackage{hyperref}
	\usepackage{graphicx}
	\usepackage{subcaption}
	\usepackage{sectsty}
	\usepackage{csquotes}
	\usepackage{tikz}
	\usepackage{tikz-cd}
	\usepackage{pst-node}
	\usepackage{dsfont}
	\usepackage{framed}
	\usepackage[all]{xy}
	
	\newcommand{\op}{\operatorname}
	\newcommand{\C}{\mathbb{C}}
	
	\newcommand{\R}{\mathbb{R}}
	
	\newcommand{\Z}{\mathbb{Z}}
	
	\newcommand{\g}{\mathfrak{g}}

	\newcommand{\into}{\hookrightarrow}

	\renewcommand{\Im}{\op{Im}}

	\sectionfont{\fontsize{12}{15}\selectfont}
	\subsectionfont{\fontsize{11}{15}\selectfont}
	\DeclareMathOperator{\Tr}{Tr}

	\DeclareMathOperator{\Hom}{Hom}
	
	\numberwithin{equation}{section}
	
	\newtheorem{defn}{Definition}[section]

	\newtheorem{lem}{Lemma}[section]
	
	\newtheorem{eg}{Example}[section]
	\newtheorem*{remark}{Remark}

	\title{Homotopy Algebras in Higher Spin Theory}
	
	\author{Si Li and Keyou Zeng}
	
	\date{\vspace{-3ex}}
	
	\newcommand{\Addresses}{{
	  \bigskip
	  \footnotesize
	
	  Si Li, \textsc{Department of Mathematical Sciences and Yau Mathematical Sciences Center (YMSC), Tsinghua University,
	    Beijing 100084, China}\par\nopagebreak
	  \textit{E-mail address}: \texttt{sili@mail.tsinghua.edu.cn}
	
	  \medskip
	
	  Keyou Zeng, \textsc{Perimeter Institute for Theoretical Physics,
	    Waterloo, Ontario, Canada, N2L 2Y5}\par\nopagebreak
	  \textit{E-mail address}: \texttt{kzeng@perimeterinstitute.ca}
	
	}}
	
	\def\curved{\tikz[baseline=.1ex]{
			\draw[ ->] (0,0.32) arc (30:330:0.5);}
	}
	\def\anticurved{\tikz[baseline=.1ex]{
			\draw[ ->] (0,0.32) arc (150:-150:0.5);}
	}
	
\begin{document}
		\maketitle
		\begin{abstract}
			Motivated by string field theory, we explore various algebraic aspects of higher spin theory and Vasiliev equation in terms of homotopy algebras. We present a systematic study of unfolded formulation developed for the higher spin equation in terms of the Maurer-Cartan equation associated to differential forms  valued in $L_\infty$-algebras.  The elimination of auxiliary variables of Vasiliev equation is analyzed through homological perturbation theory. This leads to a closed combinatorial graph formula for all the vertices of higher spin equations in the unfolded formulation. We also discover a topological quantum mechanics model whose correlation functions give  deformed higher spin vertices at first order.
		\end{abstract}
		\tableofcontents

		\section{Introduction}
		
		Field theory for point particle is usually concerned with Lie algebras and their representations, while string theory leads to generalization of Lie algebras and various kind of algebraic structures to the so called homotopy algebras. It was  known \cite{Zwiebach:1992ie,Zwiebach:1997fe} that for classical closed string field theory, its BRST operator together with its interaction vertices give us an $L_\infty$ structure (homotopy Lie structure), and classical open string field theory has an $A_\infty$ structure (homotopy associative structure). Maurer-Cartan elements on the closed string sector represent solutions of the equations of motion of closed string field theory - classical closed string backgrounds. Similarly, Maurer Cartan elements on the open string sector define a consistent classical open string field theory. The interplay is two-fold. Ideas from string theory have been a source of significant inspirations into the theory of homotopy algebras \cite{Kajiura:2004xu}. Explicit solutions of string field vertices in certain topological sectors are known and widely applied to geometry and topology: Chern-Simons type theory captures open topological string fields \cite{Witten:1992fb}; Kodaira-Spencer type gravity describes B-twisted closed topological string field theory \cite{Bershadsky:1993cx,Costello:2012cy} which is deeply related to Hodge theory. 
		
		Motivated by string field theory,  we use homotopy algebras to study higher spin field theory, a version of higher gauge theory that describes an infinite tower of massless fields of higher spins. String theory and higher spin theory are closely related. Massive higher spin excitations show up in string theory spectrum, and there are certain limits in which all these masses become negligible and the string spectrum looks like an infinite collection of massless spinning particles for every higher spin \cite{Bonelli:2003kh,Sagnotti:2003qa}. One efficient approach to higher spin theory is the unfolded formulation \cite{Vasiliev:1988sa}. The unfolded equation is originally formulated in terms of free differential algebras, and its relation with $L_\infty$ algebra is spelled out in \cite{Vasiliev:2005zu}. This is the point of view we will present for a systematic study using homotopy algebras.  As we will explain, the unfolded equation is precisely the Maurer-Cartan equation associated to differential forms  valued in certain $L_\infty$-algebras. The consistency condition is guaranteed by the $L_\infty$ structure. The higher brackets describe higher spin interactions, and solutions to the Maurer-Cartan equation give us classical higher spin backgrounds. This closely resembles the situation in string field theory.
		
		The problem of constructing field theories describing consistent propagation and interactions of higher spin fields has a long history and is highly non-trivial. This task is believed intractable in Minkovski space-time, due to Coleman-Mandula no go theorem and its generalizations which claim that symmetries of S-matrix in a non-trivial (interacting) field theory in a flat space-time can only have sufficiently low spins. On the other hand, $AdS$ space surpasses the no go theorem \cite{Bekaert:2010hw} and nontrivial interacting higher spin gauge theories have been constructed via Vasiliev's system in $AdS_4$ and its generalization in various dimensions \cite{Vasiliev:1990en,Vasiliev:1990vu,Vasiliev:1992av}. Vasiliev theory is constructed at the level of equations of motion, based on the unfolded approach and introduction of auxiliary variables. Our first goal in this paper is to explore and clarify various algebraic structures behind Vasiliev equation, and we find homotopy algebras fit into business.   
		
		Our formulation lies on an important tool in homotopy algebra called the Minimal Model Theorem. Briefly speaking, it states that for a given $A_\infty$ algebra, there exists an $A_\infty$ structure on its (co)homology which is $A_\infty$ quasi-isomorphic to the original one. We will present a concrete approach to Minimal Model Theorem using	 
		homological perturbation theory, 
		which is a computational tool for transferring differentials along homotopy equivalences of chain complexes. It enable us to obtain new homotopy algebraic structure on the (co)homology of the original algebra. In certain situations, the algebras under consideration are related to physical theory, then the homological perturbation has physical interpretation. Generally speaking, two complexes related by a quasi-isomorphism correspond to two theories related by renormalization or addition and elimination of auxiliary fields. There homological perturbation is the same as integrating out high energy modes, and various formulas that appear in homological perturbation have physical interpretation -- summing over Feynman diagrams. In this case Minimal Model Theorem is about transferring a simple algebraic structure for fields at UV to a homotopy one on zero modes which organizes the structure of all higher effective vertices.

		 In our applications to higher spin theory and Vasiliev equation, we consider the process of elimination of the auxiliary variable of Vasiliev equation. The DGLA (or $L_\infty$) structure of Vasiliev equation is transferred to an $L_\infty$ structure on its homology, which is the physical degree of freedom of higher spin equations. The set of higher brackets computed through homological perturbation gives us consistent interacting vertices of higher spin fields. Similar reduction process is analyzed in \cite{Sezgin:2000hr,Sezgin:2002ru} and is referred to as curvature expansion therein. Using homological perturbation, we obtain a closed combinatorial graph formula for all the vertices of higher spin equations (formula \eqref{tree_formula_2}). 
		 	
		Another possible way to construct nonlinear higher spin equation in the unfolded approach is to construct the vertex by hands and check the consistency equation order by order \cite{Vasiliev:1988sa}. This can be viewed as a deformation problem. Generally, the first order deformation of an algebra is controlled by its Hochschild cohomology. Alternately, the consistency condition for the first order vertex gives us Hochschild cocycle condition \cite{Sharapov:2017yde}. In higher spin theory, this is closely related to the Hochschild cohomology of Weyl algebras which was explicitly computed in \cite{Feigin}. We illustrate a connection between the deformed vertex and a topological quantum mechanics model \cite{Si-index}, where the Hochschild cocycle condition is implemented through the quantum master equation. This sheds light on the first quantized nature of Vasiliev theory.
		
		One important aspect of higher spin theory is its fundamental role in AdS/CFT correspondence. Vasiliev's higher spin gauge theory in $AdS_4$ is conjectured to be holographically dual to $O(N)$
		Vector Models \cite{Klebanov:2002ja}, and non-trivial checks of the duality appear in \cite{Giombi:2009wh,Giombi:2010vg}. This duality may be regarded as the simplest non-trivial example of AdS/CFT correspondence. There are recent works \cite{Costello:2017fbo,Costello:2016mgj} suggesting relations between AdS/CFT correspondence and an algebraic operation called \emph{Koszul duality}. Our results provide mathematical preliminaries to study the structural aspect of higher spin theory. We hope to address its link with AdS/CFT and Koszul duality in our later work.
		
		The organization of this paper is as follows. Section 2 gives an introduction to the $A_\infty$ and $L_\infty$ algebra. Section 3 introduces homological perturbation theory and we illustrate its physics meaning via the example of Chern-Simons theory. In section 4, we introduce higher spin theory and Vasiliev equation, and apply homological perturbation theory to analyze Vasiliev equation. In section 5, we introduce a topological quantum mechanics model whose correlation function give us interaction vertices in higher spin equation.

		\section{Homotopy algebras} In this section we introduce basic examples of homotopy algebras: the $A_\infty$-algebra (homotopy associative algebra) and the $L_\infty$-algebra (homotopy Lie algebra). We discuss the associated Maurer-Cartan equations and explain how Maurer-Cartan elements can be used to twist homotopy algebras. Physically, this corresponds to expand the theory around a background, which will play an important role in our applications to higher spin theory and Vasiliev equation.
		\vskip 0.5 cm
		\noindent \textbf{Conventions}
		\newline
		
		We will mostly work with $\mathbb{Z}$-graded $\C$-vector space
		$$
		V = \bigoplus\limits_{n\in \mathbb{Z}}V_n.
		$$
		The grading $n$ is related to the ghost number in physics. The degree of an element $v \in V_n$ is denoted by $|v| = n$, and such a $v$ is called a homogeneous element.
		
		For $V$ and $W$ two graded vector spaces,  $V\otimes W$ denotes tensor product over $\C$ and  $\Hom(V,W)$ denotes $\C$-linear maps (continuous with appropriate topology that we will always be sloppy in our discussion). They are all graded vector spaces where
		\begin{equation*}
			(V\otimes W)_n = \bigoplus_{i+j = n} V_{i}\otimes W_j, \quad \Hom(V,W)_{n} = \bigoplus_{i}\Hom(V_i,W_{i+n}).
		\end{equation*}
		In particular, the linear dual $V^*$ is a graded vector space with $(V^*)_m = \Hom(V_{-m} , \mathbb{C})$.
		
		We denote the Koszul sign braiding on tensor products to be
		\begin{equation*}
			\begin{aligned}
				\tau_{V,W}:V\otimes W &\to W\otimes V\\
				v\otimes w &\mapsto (-1)^{|v||w|} w\otimes v
			\end{aligned}
		\end{equation*}
		In the above formula  $v$ and $w$ are supposed to be homogeneous. Then $\tau_{V,W}$ is extended to general elements by linearity. This sign rule appears naturally in physics from interchanging bosons or fermions. In general we can consider interchanging $n$ particles, then the above sign rule induces naturally a sign rule for the action of symmetric group $S_n$ on the $n$-th tensor product $V^{\otimes n}$
		\begin{equation*}
			v_1\otimes v_2\otimes \cdots \otimes v_n \to \epsilon(\sigma,v)v_{\sigma(1)}\otimes v_{\sigma(2)} \otimes \dots v_{\sigma(n)}
		\end{equation*}
		where $\epsilon(\sigma,v)$ is called the Koszul sign.
		
		This sign rule has very important consequences. It explains, for instance, the sign rule for tensor products of maps. Given graded vector spaces $V,W$, we have the following chain of linear maps:
		\begin{equation*}
			(V^*\otimes W^*)\otimes(V\otimes W) \overset{\tau_{W^*,V}}{\cong} V^*\otimes V \otimes W^*\otimes W \overset{\varphi_V\otimes \varphi_W}{\to}  \C\otimes \C \cong \mathbb{C}
		\end{equation*}
		where $\varphi_V: V^*\otimes V\to \C$ is the natural pairing. This gives us the following formula for $f\in V^*,g \in W^*,v\in V, w\in W$:
		\begin{equation*}
			(f\otimes g)(v\otimes w) = (-1)^{|v||g|}f(v)\otimes g(w).
		\end{equation*}
		Similarly for  $f,f'\in
		\Hom(V, V),g,g' \in \Hom(W, W)$:
		\begin{equation*}
			(f\otimes g)\circ(f'\otimes g') = (-1)^{|f'||g|}(f\circ f')\otimes (g\circ g')
		\end{equation*}
		where $\circ$ is the composition of linear maps.
		
		\subsection{$A_\infty$-algebra}
		In this subsection we briefly introduce $A_\infty$ algebra. See e.g. \cite{Keller} for more details.
		\begin{defn}[$A_\infty$-algebra] Let $A$ be a $\mathbb{Z}$ graded vector space $A= \bigoplus_n A_n$. An $A_\infty$ structure on $A$ is a collection of linear maps $m_k: A^{\otimes k} \to A$ of degree $2 - k$ ($k \geq 1$) that satisfy the identities
			\begin{equation}\label{A_inf_rel}
				\sum_{r = 0}^{n-1}\sum_{k = 1}^{n - r} (-1)^{rk + (n-k-r)}m_{n - k + 1}\circ(\mathds{1}^{\otimes r}\otimes m_k \otimes \mathds{1}^{\otimes (n - k - r)}) = 0,\quad \;n\geq 1
			\end{equation}
		\end{defn}
		
		Let us analyze the defining relations for small values of $n$:
		\begin{enumerate}
			\item[1)] $n = 1$. The relation is $m_1\circ m_1 = 0$, which means that $m_1$ is a differential on $A$.
			\item[2)] $n = 2$. We have $m_1\circ m_2 = m_2\circ(\mathds{1}\otimes m_1 + m_1\otimes\mathds{1})$, or equivalently
			\begin{equation*}
				m_1(m_2(x_1,x_2)) = m_2(m_1(x_1),x_2) + (-1)^{|x_1|}m_2(x_1,m_1(x_2))
			\end{equation*}
			which says that $m_1$ is a derivation with respect to the product on $A$ defined by $m_2$.
			\item[3)] $n = 3$. The relation yields
			$$
			m_2\circ(m_2\otimes \mathds{1} - \mathds{1}\otimes m_2) = m_1\circ m_3 + m_3(m_1\otimes \mathds{1}^{\otimes 2} + \mathds{1}\otimes m_1\otimes \mathds{1} + \mathds{1}^{\otimes 2}\otimes m_1).
			$$ This says that $m_2$ is an associative product up to homotopy given by $m_3$.
		\end{enumerate}
		
		\paragraph{Tree description.}
		We pictorially represent each product $m_k$ as
		$$
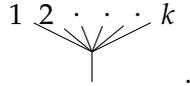
\begin{tikzpicture}[grow' = up]
		\tikzstyle{level 1}=[level distance = 4mm]
		\tikzstyle{level 2}=[sibling distance=4mm,level distance = 5mm]
		\coordinate
		child {
			child {node {$1$}} child {node {$2$}} child {node {$.$}} child {node {$.$}} child {node {$.$}} child {node {$k$}}
		};
		\end{tikzpicture}.
		$$
		We define a differential $\partial$ on $\Hom(A^{\otimes n},A)$ by
		$$
		\partial(f) = m_1\circ f - (-1)^{|f|}\sum_{r= 0}^{n-1}f\circ (\mathds{1}^{\otimes r}\otimes m_1 \otimes \mathds{1}^{\otimes (n - 1 - r)}), \quad f \in \Hom(A^{\otimes n},A).
		$$
		Then the $A_\infty$ relation can be represented as
		\begin{center}
			\begin{tikzpicture}[grow' = up]
			\tikzstyle{level 1}=[level distance = 4mm]
			\tikzstyle{level 2}=[sibling distance=4.5mm,level distance = 5mm]
			\tikzstyle{level 3}=[sibling distance=4.5mm,level distance = 5mm]
			\tikzstyle{level 4}=[sibling distance=4.5mm,level distance = 5mm]
			\coordinate
			child {
				child {node {$1$}} child {node {$\dots$}} child {node {$.$}} child {node {$j$} child {child {node {$1$}}  child {node {$\dots$}} child {node {$l$}} }} child {node {$\dots$}}  child {node {$k$}}
			};
			\node at (-3,1) {$\sum\limits_{\substack{k+l=n + 1 \\ k,l\geq2}}\sum_{1\leq j \leq k}\pm $};
			\node at (2,1) {$=$};
			\node at (2.5,1) {$\partial\left(\right.$};
			\end{tikzpicture}
			\begin{tikzpicture}[grow' = up]
			\tikzstyle{level 1}=[level distance = 5mm]
			\tikzstyle{level 2}=[sibling distance=8mm,level distance = 6mm]
			\coordinate
			child {
				child {node {$1$}} child {node {$2$}} child {node {$...$}} child {node {$n$}}
			};
			\node at (1.4,1) {$\left.\right)$};
			\end{tikzpicture}
		\end{center}
		The general structure for an $A_\infty$ algebra is that the maps satisfy versions of certain associativity identities up to homotopy.

		~
		\begin{eg}$\empty$
			\begin{enumerate}
				\item An $A_\infty$ algebra with $m_k = 0$ for $k \geq 3$ is called a differential graded algebra (DGA). We usually denote a DGA by $(A,d,\cdot)$.
				\item A commutative differential graded algebra (CDGA ) is a differential graded algebra $(A,d,\cdot)$ with supercommutative product: for any two homogeneous elements $x,y$ 
				\begin{equation*}
					x\cdot y = (-1)^{|x||y|}y\cdot x
				\end{equation*}
				
				\item The de Rham complex of differential forms on a smooth manifold, equipped with the de Rham differential and the wedge product form a CDGA $(\Omega^{\bullet}(X),d,\,\wedge\,)$.
			\end{enumerate}
		\end{eg}
		
		\subsection{$L_\infty$-algebra}
		An $L_\infty$-structure is a natural generalization of a graded Lie algebra that includes an infinite number of skew-symmetric multi-linear ``brackets" satisfying an infinite number of generalized Jacobi identities. It also appears many places in physics. For example, Zwiebach found that the $n$-point functions equip the BRST complex in the closed string field theory with such a structure \cite{Zwiebach:1992ie}. For more detailed introduction to $L_\infty$ algebra, see e.g \cite{Lada:1992wc}. Our sign convention follows \cite{Getzler}.
		
		\begin{defn}[$L_\infty$-algebra] Let $\mathfrak{g}$ be a $\mathbb{Z}$ graded vector space $\mathfrak{g} = \bigoplus_n \mathfrak{g}_n$. An $L_\infty$ structure on $\mathfrak{g}$ is a collection of multi-linear maps $l_k: \mathfrak{g}^{\otimes k} \to \mathfrak{g}$ of degree $2 - k$ ($k\geq1$) that are graded skew-symmetric
			\begin{equation*}
				l_n(x_{\sigma(1)},\dots,x_{\sigma(n)}) = (-1)^\sigma\epsilon(\sigma,x)l_n(x_1,\dots,x_n)
			\end{equation*}
			and satisfy the following identities
			\begin{equation*}
				\sum_{k = 1}^n(-1)^{k}\sum_{\sigma \in \textbf{Unsh}(k,n-k)} (-1)^\sigma\epsilon(\sigma,x)l_{n - k + 1}(l_k(x_{\sigma(1)},\dots,x_{\sigma(k)}),x_{\sigma(k+1)},\dots,x_{\sigma(n)}) = 0
			\end{equation*}
			where $\epsilon(\sigma,x)$ is the Koszul sign of the permutation $\sigma$, and $\textbf{Unsh}(k,n-k)$ is a subset of the permutation group called $(k,n-k)$-unshuffle. It is the set of permutation $\sigma$ such that
			\begin{equation*}
				\sigma(1) < \sigma(2) < \dots < \sigma(k),\;\;\sigma(k+1)< \dots < \sigma(n)
			\end{equation*}
			The Cardinality of $\textbf{Unsh}(k,n-k)$ is $\binom{n}{k}$
		\end{defn}
		Let us analyze the defining relations for small values of $n$:
		\begin{enumerate}
			\item[1)] $n = 1$. The relation is $l_1\circ l_1 = 0$, which means that $l_1$ is a differential on $\mathfrak{g}$.
			\item[2)] $n = 2$. We have
			\begin{equation*}
				l_1(l_2(x_1,x_2)) = l_2(l_1(x_1),x_2) + (-1)^{|x_1|}l_2(x_1,l_1(x_2))
			\end{equation*}
			which says that $l_1$ is a derivation with respect to the binary map $l_2$.
			\item[3)] $n = 3$. The relations yields
			\begin{equation*}
				\begin{aligned}
					& l_2(l_2(x_1,x_2),x_3) + (-1)^{(|x_1|+|x_2|)|x_3|}l_2(l_2(x_3,x_1),x_2)  + (-1)^{(|x_2|+|x_3|)|x_1|}l_2(l_2(x_2,x_3),x_1) \\
					&= l_1l_3(x_1,x_2,x_3) + l_3(l_1(x_1),x_2,x_3) + (-1)^{|x_1|}l_3(x_1,l_1(x_2),x_3) + (-1)^{|x_1| + |x_2|}l_3(x_1,x_2,l_1(x_3))
				\end{aligned}
			\end{equation*}
			which says that $l_2$ satisfies Jacobi identities up to homotopy given by $l_3$.
		\end{enumerate}
		
		\begin{eg}
			~
			\begin{enumerate}
				\item An $L_\infty$-algebra with $l_k = 0$ for $k \geq 3$ is called a differential graded Lie algebra (DGLA). We usually denote a DGLA by $(\mathfrak{g},d,[-,-])$.
				\item For $\mathfrak{g}$ an $L_\infty$ algebra, $(A,d_A)$ a commutative differential graded algebra, the tensor product $A\otimes \mathfrak{g}$ naturally inherits the structure of an $L_\infty$ algebra. Denote $\{l_k: \mathfrak{g}^{\otimes k} \to \mathfrak{g},\;k\geq1\}$ the $L_\infty$ structure on $\mathfrak{g}$, then the $L_\infty$ structure on $A\otimes \mathfrak{g}$ is given by
				\begin{equation*}
					\begin{aligned}
						l_1'(a \otimes x) &= d_Aa \otimes x +    + (-1)^{|a|}a \otimes l_1(x)\\
						l'_k(a_1 \otimes x_1 ,a_n \otimes x_n)& = (-1)^{\sum_{i<j}|x_{i}||a_j|}a_1\cdot a_2\cdots a_n \otimes l_k(x_1,\dots,x_n), \quad k>1
					\end{aligned}
				\end{equation*}
				where $\sum_{i<j}|x_{i}||a_j|$ comes from Koszul sign.
				
				In particular, when $A = \Omega^{\bullet}(X)$, $\Omega^{\bullet}(X)\otimes \mathfrak{g}$ naturally inherits an $L_\infty$ structure.
				\item For an associative algebra $V$ with product $a\otimes b \mapsto a\cdot b$, the commutator of this product $[x_1,x_2] = x_1\cdot x_2 - (-1)^{|x_1||x_2|}x_2\cdot x_1$ defines a Lie algebra structure on $V$. Similarly, if $\{m_k: V^{\otimes k} \to V,\;k \geq 1\}$ is a $A_\infty$ structure on $V$, then the collection $\{l_k: V^{\otimes k} \to  V,\;k\geq1\}$ defined by anti-symmetrization
				\begin{equation*}
					l_n(x_1,\dots,x_n) = \sum_{\sigma}(-1)^{\sigma}\epsilon(\sigma,x)m_n(x_{\sigma(1)},\dots,x_{\sigma(n)})
				\end{equation*}
				defines an $L_\infty$ structure.
			\end{enumerate}
		\end{eg}

		\subsection{Algebras encoded in coderivations}
		There are equivalent descriptions of $A_\infty$ and $L_\infty$ algebras in terms of degree one coderivations on the free coalgebras $T^c(V[1])$, $S^c(V[1])$. The defining relation is compactly encoded in the single equation saying that the coderivation squares to zero. Roughly speaking, the geometric counterpart of an $L_\infty$ algebra is a formal pointed manifold equipped with a cohomological vector field \cite{Kontsevich:1997vb}. Similarly an $A_\infty$ is the same as a cohomological vector field on a non-commutative formal manifold. In this section we briefly review this construction. See e.g. \cite{Getzler:1990} for more detail.
		
		Naively, "coalgebras are the dual of algebras".  Given a (finite dimensional) algebra $A$ with an associative product $\mu : A\otimes A \to A$, its dual gives us a map $\mu^*: A^* \to A^*\otimes A^* $. The associativity of $\mu$ implies that $\mu^*$ satisfy $(\mu^*\otimes 1)\circ\mu^* = ( 1 \otimes \mu^*)\circ\mu^* $. This motivates the general definition of coalgebra:
		\begin{defn}$\empty$
			
			\begin{enumerate}
				\item[(1)] 	A (graded) coalgebra is a (graded) module $C$ with a comultiplication $\Delta : C\to C\otimes C$ of degree 0 such that the following diagram commutes (co-associativity)
				
				\[
				\xymatrix{
					C \ar[r]^-{\Delta}\ar[d]_-{\Delta} & C\otimes C \ar[d]^-{\Delta\otimes 1}\\
					C\otimes C\ar[r]_-{1\otimes \Delta} & C\otimes C\otimes C
				}\,.
				\]
				\item[(2)] A coderivation on a coalgebra is a map $ L : C \to C$ such that the following diagram commutes (co-Leibniz rule):
				\[\xymatrixcolsep{5pc}
				\xymatrix{
					C \ar[r]^-{L}\ar[d]_-{\Delta} & C \ar[d]^-{\Delta\otimes 1}\\
					C\otimes C\ar[r]_-{1\otimes L + L \otimes 1} & C\otimes C
				}\,.
				\]
				\item[(3)] A differential graded coalgebra is a graded coalgebra with a coderivation $b: C\to C$ of degree 1 such that $b^2 = 0$.
			\end{enumerate}
		\end{defn}
		The basic example of a graded coalgebra is the tensor coalgebra of a graded module
		\begin{equation*}
			T^c(V) = \bigoplus_{n = 0}^\infty V^{\otimes n}
		\end{equation*}
		with comultiplication:
		\begin{equation}
			\Delta(v_1\otimes\cdots\otimes v_n) = \sum_{i = 0}^{n}(v_1\otimes \cdots \otimes v_i)\otimes (v_{i+1}\otimes \cdots \otimes v_n)
		\end{equation}
		We will frequently work with the reduced tensor coalgebra
		$$
		\bar T^c(V) = \bigoplus_{n = 1}^\infty V^{\otimes n}
		$$
		with the reduced comultiplication
		\begin{equation}
			\bar \Delta(v_1\otimes\cdots\otimes v_n) = \sum_{i = 1}^{n-1}(v_1\otimes \cdots \otimes v_i)\otimes (v_{i+1}\otimes \cdots \otimes v_n).
		\end{equation}

		For a tensor algebra, a derivation is completely specified by its action on the generator: the Leibniz rule tells us how to extend
		the derivation on tensors. A dual statement is also true: a coderivation is completely specified by the $V$ component of its action. That is, we have the following identification
		\begin{equation}
			\text{Coder}(\bar T^c(V)) \cong \text{Hom}(\bar T^c(V),V)
		\end{equation}
		To see this, denote
		$$
		\text{proj}_V: \bar T^c(V)\to V
		$$
		the natural projection and
		$$
		j_{V^{\otimes n}}: V^{\otimes n}\into \bar T^c(V)
		$$
		the natural inclusion. On one hand, a coderivation $L$ determines a map
		\begin{equation*}
			\text{proj}_V \circ L \in \text{Hom}(\bar T^c(V),V)
		\end{equation*}
		or a set of maps $L_k = \text{proj}_V \circ L \circ j_{V^{\otimes n}} \in \Hom(V^{\otimes k},V),k\geq 1$. On the other hand, given a set of maps $\{L_k\in Hom(V^{\otimes k},V),k\geq 1\}$ or an element in $\text{Hom}(\bar T^c(V),V)$, it determines
		\begin{equation}
			L = \sum_{i\geq 1}^n\sum_{j = 0}^{n-i} \mathds{1}^{\otimes j}\otimes L_{i}\otimes \mathds{1}^{n - i - j}
		\end{equation}
		as a coderivation, and $L$ defined above satisfies $L_i = \text{proj}_V \circ L \circ j_{V^{\otimes i}}$.
		
		If $b$ is a coderivation of degree $1$ on $\bar T^c(V)$ with components $b_n: V^{\otimes n} \to V$, then its square is a coderivation of degree $2$ with components
		\begin{equation}
			(b^2)_n = \sum_{i+j = n+1}\sum_{k = 0}^{n - j} b_i\circ(\mathds{1}^{\otimes k}\otimes b_{j}\otimes \mathds{1}^{n - k - j})
		\end{equation}
		For $b$ to be a differential we must have all the $(b^2)_n$ vanish. We see that the resulting relation is similar to the defining relation for an $A_\infty$ algebra except for the sign difference. This is encoded in the degree shifting as follows.
		
		For $V$ a $\mathbb{Z}$ graded vector space, we denote $V[n]$ the degree $n$-shifted space such that
		\begin{equation}
			V[n]_{m} := V_{n+m}
		\end{equation}
		We also use the notation of suspension $sV$ defined by the tensor product
		\begin{equation}
			sV := \mathbb{C}s\otimes V
		\end{equation}
		where $\mathbb{C}s$ is the one-dimensional graded vector space spanned by $s$ with $|s| = -1$. In particular, $(sV)_{m} = V_{m+1}$, or $sV = V[1]$. We can also regard $s$ as a degree $-1$ linear map $s: V \to V[1]$. For a homogeneous $a \in V$, we have $sa \in V[1]$ and $|sa| = |a| - 1$.
		
		Similarly, let $\mathbb{C}s^{-1}$ be the graded vector space spanned by $s^{-1}$ sitting at degree $1$. We define the desuspension of the graded space $V$ by
		$$
		s^{-1}V: = \mathbb{C}s^{-1} \otimes V.
		$$
		In particular $(s^{-1}V)_{m} = V_{m-1}$. We can also regard $s^{-1}$ as a degree $1$ linear map, such that $s^{-1}s = ss^{-1} = 1$. The desuspension map induces maps $(s^{-1})^{\otimes k}: (V[1])^{\otimes k} \to V^{\otimes k} $.
		
		For a linear map $m_k: V^{\otimes k } \to V $, it can be identified with a linear map $b_k: (V[1])^{\otimes k} \to V[1]$ defined by $b_k = s\circ m_k\circ (s^{-1})^{\otimes k}$, or by the following commutative digram
		
		\[
		\xymatrix{
			(V[1])^{\otimes k}  \ar[r]^-{b_k}\ar[d]_-{(s^{-1})^{\otimes k}} & V[1] \ar[d]^-{s^{-1}}\\
			V^{\otimes k}\ar[r]_-{m_k} & V
		}\,.
		\] 
		The Koszul sign convention gives
		\begin{equation}
			\begin{aligned}
				b_k(sa_1\otimes \dots sa_k) &= s m_k (s^{-1})^{\otimes k}(sa_1\otimes \dots sa_k)\\
				&  = (-1)^{\sum_{i=1}^{k-1}(k-i)|a_i| - k(k-1)/2}s m_k(a_1\otimes\dots a_k).
			\end{aligned}
		\end{equation}
		Note that $s^{\otimes k} (s^{-1})^{\otimes k} = (-1)^{k(k-1)/2}$. Therefore
		$$
		b_k = s\circ m_k\circ (s^{-1})^{\otimes k} \Longleftrightarrow m_k = (-1)^{k(k-1)/2}s^{-1}\circ m_k\circ s^{\otimes k}.
		$$

		For maps $\{m_k\}_{k\geq 1}$ satisfying the $A_\infty$ relation \eqref{A_inf_rel}, we consider maps $b_k = s\circ m_k\circ (s^{-1})^{\otimes k}$. It defines a coderivation whose square is
		\begin{equation*}
			\begin{aligned}
				&\sum_{r = 0}^{n-1}\sum_{k = 1}^{n - r} b_{n - k + 1}\circ(\mathds{1}^{\otimes r}\otimes b_k \otimes \mathds{1}^{\otimes (n - k - r)})\\
				= &\sum_{r = 0}^{n-1}\sum_{k = 1}^{n - r} s\circ m_{n - k + 1}\circ(s^{-1})^{\otimes n-k+1}\circ(\mathds{1}^{\otimes r}\otimes s\circ m_k\circ (s^{-1})^{\otimes k} \otimes \mathds{1}^{\otimes (n - k - r)}) \\
				=& \sum_{r = 0}^{n-1}\sum_{k = 1}^{n - r} (-1)^{n-k-r}s\circ m_{n - k + 1}\circ ((s^{-1})^{\otimes r}\otimes m_k\circ s^{\otimes k} \otimes (s^{-1})^{\otimes (n - k - r)})\\
				=&\sum_{r = 0}^{n-1}\sum_{k = 1}^{n - r}(-1)^{kr+ (n - k -r)} s\circ m_{n - k + 1}\circ(\mathds{1}^{\otimes r}\otimes m_k\otimes \mathds{1}^{\otimes (n - k - r)})\circ (s^{-1})^{\otimes n}\\
				=&0
			\end{aligned}
		\end{equation*}
		
		This shows that the following data are equivalent
		\begin{framed}
			\begin{itemize}
				\item  A collection of linear maps $m_k: A^{\otimes k} \to A$ of degree $2 - k$ satisfying $A_\infty$ relation.
				\item A degree $1$ coderivation $b$ on $\bar T^c(A[1])$ satisfying $b^2 = 0$.
			\end{itemize}
		\end{framed}
		\begin{remark}
			The coalgebra definition for $A_\infty$ algebra is unambiguous. However, there are two conventions of the $A_\infty$ relation for $m_k,\; k\geq 1$, one is obtain by defining $b_k = s\circ m_k\circ (s^{-1})^{\otimes k}$ as what we use, the other is through $m_k = s^{-1}\circ b_k\circ s^{\otimes k}$. These two convention lead to different $m_k$ that differ by a sign $s^{\otimes k} (s^{-1})^{\otimes k} = k(k-1)/2$. We take our convention so that the MC equation is simple without extra $\pm$ sign.
		\end{remark}
		
		There is a similar construction for the $L_\infty$ algebra. Instead of the tensor coalgebra, we consider the cocommutative coalgebra $S^{c}(V)$ and its reduced version $\bar S^c(V)$ where
		\begin{equation*}
			S^c(V) = \bigoplus_{n = 0}^\infty S^n(V) :=T^c(V)/I_S, \quad \bar S^c(V) = \bigoplus_{n = 1}^\infty S^n(V) :=\bar T^c(V)/I_S.
		\end{equation*}
		Here $I_S$ is the subspace generated by vectors of the form.
		\begin{equation*}
			v_1\otimes\cdots \otimes v_n - \epsilon(\sigma,v)v_{\sigma(1)}\otimes \cdots \otimes v_{\sigma(n)}.
		\end{equation*}
		In other words, $S^c(V)$ is the graded symmetric tensors, i.e., the quotient of $T^c(V)$ by the graded permutations. Let
		$$
		\pi_S:V^{\otimes n}\to S^n(V)
		$$
		be the natural quotient map and we denote
		\begin{equation*}
			v_1\odot\cdots\odot v_n := \pi_S(v_1\otimes\cdots \otimes v_n).
		\end{equation*}
		We define the coproduct $\bar \Delta: \bar S^c(V) \to \bar S^c(V)\otimes \bar S^c(V)$ by
		\begin{equation}
			\bar \Delta(v_1\odot\cdots\odot v_n) = \sum_{i = 1}^{n-1}\sum_{\sigma \in \textbf{Unsh}(i,n-i)}\epsilon(\sigma,v)(v_{\sigma(1)}\odot\cdots\odot v_{\sigma(i)})\otimes(v_{\sigma(i+1)}\odot\cdots\odot v_{\sigma(n)})
		\end{equation}
		Then the following data are equivalent
		\begin{framed}
			\begin{itemize}
				\item  A collection of linear maps $l_k: \mathfrak{g}^{\otimes k} \to \mathfrak{g}$ of degree $2 - k$  satisfying $L_\infty$ relation.
				\item  A degree $1$ coderivation $Q$ on $\bar S^c(\mathfrak{g}[1])$ satisfying $Q^2 = 0$.
			\end{itemize}
		\end{framed}
		The abstract definition via codifferential on coalgebra enables us to establish powerful techniques like homological perturbation theory that will be introduced in the next section.
		
		\subsection{Maurer-Cartan equation}
		In this subsection, we discuss the Maurer-Cartan equation associated to an $A_\infty$ or $L_\infty$ algebra. We will show that given a Maurer-Cartan element, we can twist the original $A_\infty$ (or $L_\infty$) algebra to get a new $A_\infty$ (or $L_\infty$) algebra. Physically, this corresponds to expand the theory around a background solving the equation of motions.
		
		\paragraph{Maurer-Cartan equation.}
		Given an $A_\infty$-algebra $(A,m_1,m_2,\dots)$, a Maurer-Cartan element is a degree $1$ element $\alpha \in A_1$ satisfying
		\begin{equation}\label{MC-equation}
			\sum_{k\geq 1}m_k(\alpha,\dots,\alpha) = 0.
		\end{equation}
		The above equation is called Maurer-Cartan equation. The set of Maurer-Cartan elements is denoted by $MC(A)$. Similarly, given an $L_\infty$-algebra $\mathfrak{g}$, a Maurer-Cartan element is a degree $1$ element $\alpha \in \mathfrak{g}_1$ satisfying
		\begin{equation}\label{MC-equation2}
			\sum_{k\geq 1}\frac{1}{k!}l_k(\alpha,\dots,\alpha) = 0.
		\end{equation}
		The set of Maurer-Cartan elements is still denoted by $MC(\mathfrak{g})$. Note that if $\mathfrak{g}$ comes from anti-symmetrization of an $A_\infty$ algebra $A$, then the Maurer-Cartan equation for $A$ is equivalent to the MC equation for $\mathfrak{g}$.
		
		\begin{remark}
			Strictly speaking, the Maurer-Cartan equation is defined as a functor on local algebras: let $R$ be a finite dimensional $\C$-algebra and $R_+$ its maximal ideal. There exists $N>0$ such that $R_+^N=0$. Let $A(R):=A\otimes R$, which is called the $R$-point of $A$ in the language of algebraic geometry. $A(R)$ is naturally an $A_\infty$ algebra by $R$-linear extension. Precisely, $\alpha$ in the Maurer-Cartan equation \eqref{MC-equation} lies in $(A\otimes R_+)_1$. Then $\alpha\otimes \cdots_{k\ \text{times}}\otimes \alpha=0$ for $k\geq N$ and \eqref{MC-equation} is a finite sum, hence well-defined. This construction is functorial in $R$, leading to the formal mathematical meaning of \eqref{MC-equation} \eqref{MC-equation2}. We will ignore this subtlety in our later discussion to simplify the presentation. Careful reader can fix the sloppy formulae.
		\end{remark}
		
		\paragraph{Twisting homotopy structures.} Maurer-Carten elements can be used to twist the original $A_\infty$ ($L_\infty$) structures to arrive at new $A_\infty$ ($L_\infty$) structures.
		
		\paragraph{$A_\infty$ case.}
		First we consider the $A_\infty$ case. Given an $A_\infty$ algebra $A$, let $b$ be the associated coderivation on $\bar T^c(A[1])$. $b$ is equivalently described by
		$$
		b_* = b_1 + b_2 + \dots = \text{proj}_{A[1]}\circ b \in \text{Hom}(\bar T^c(A[1]),A[1]).
		$$
		We naturally extend $b_*$ as a coderivation on $T^c(A[1])$ in terms of
		$$
		b_*\in \text{Hom}(\bar T^c(A[1]),A[1])\to \Hom(T^c(A[1]), A[1])=\text{Coder}(T^c(A[1])).
		$$
		
		Given an element $\Phi \in (A[1])_0$, we introduce the notation
		\begin{equation*}
			e^{\Phi} = 1+\Phi + \Phi\otimes \Phi + \dots = \sum_{k\geq 0}\Phi^{\otimes k}.
		\end{equation*}
		
		Assume $\Phi$ solves the Maurer-Carntan equation $b_*(e^\Phi)=0$, which is equivalent to
		$$
		b(e^\Phi)=e^\Phi\otimes b_*(e^\Phi)\otimes e^{\Phi}=0.
		$$
		We define new $A_\infty$ operations by
		\begin{equation}
			\boxed{b_k^\Phi(x_1\otimes \cdots\otimes x_k):=b_*(e^\Phi\otimes x_1\otimes e^{\Phi}\otimes x_2\otimes \cdots\otimes e^{\Phi}\otimes x_k\otimes e^{\Phi}). }
		\end{equation}
		It defines a new coderivation $b^\Phi$. It is a good exercise to show that Maurer-Cartan equation of $\Phi$ implies
		$$
		b^\Phi\circ b^\Phi=0.
		$$
		This shows that $(\bar T^c(A[1]), b^\Phi)$ defines a new $A_\infty$ structure. For example, the new differential $b^{\Phi}_1: A[1] \to A[1]$ is given by
		\begin{equation*}
			b^{\Phi}_1(x) = \sum_{m,n\geq 0}b_{m+n+1}(\Phi^{\otimes m}\otimes x \otimes \Phi^{\otimes n}),\; x \in A[1].
		\end{equation*}
		The twisting can be stated using the maps $m_n$ on unshifted space $A$. For $\varphi \in A_1$ a Maurer-Cartan element, the set of maps
		\begin{equation*}
			m^{\varphi}_k(x_1,\dots,x_k) := \sum_{l\geq 0 }\sum_{l_1+ \dots l_{k+1} = l} (-1)^{\sum_{i=1}^k (l -i -\sum_{j=1}^{i}l_j)(|x_i| - 1)}m_{k+l}(\varphi^{\otimes l_1}\otimes x_1 \otimes \varphi^{\otimes l_2}\otimes  \dots x_k \otimes\varphi^{\otimes l_{k+1}})
		\end{equation*}
		satisfy the $A_\infty$ relations \eqref{A_inf_rel}.

		\paragraph{$L_\infty$ case.}
		The case for $L_\infty$ algebra $(\bar S^c(\mathfrak{g}[1]),Q)$ is similar. $Q$ is a coderivation on $\bar S^c(\mathfrak{g}[1])$, which can be viewed naturally as a coderivation on $S^c(\mathfrak{g}[1])$. $Q$ is determined by
		$$
		Q_* = \text{proj}_{\mathfrak{g}[1]}\circ Q\in \Hom(\bar S^c(\mathfrak{g}[1]),\mathfrak{g}[1]).
		$$
		
		Given $\Phi \in (\mathfrak{g}[1])_0$, let
		\begin{equation*}
			e^{\Phi} = 1+\Phi + \frac{1}{2}\Phi\odot \Phi + \dots = \sum_{k\geq 0}\frac{1}{k!}\Phi^{\odot k}.
		\end{equation*}
		
		Assume $\Phi$ solves the Maurer-Carntan equation $Q_*(e^\Phi)=0$, which is equivalent to
		$$
		Q(e^\Phi)=e^\Phi\odot  Q_*(e^\Phi)\odot  e^{\Phi}=0.
		$$
		We define new $L_\infty$ operations by
		\begin{equation}
			\boxed{Q_k^\Phi(x_1\odot \cdots\odot x_k):=Q_*(e^\Phi\odot x_1\odot x_2\odot\cdots\odot x_k). }
		\end{equation}
		It defines a new coderivation $Q^\Phi$ such that
		$$
		Q^\Phi\circ Q^\Phi=0.
		$$
		This shows that $(\bar S^c(A[1]), Q^\Phi)$ defines a new $L_\infty$ structure.
		
		The twisting can be reformulated for maps $l_k,\; k \geq 1$ on unshifted space $\mathfrak{g}$. For $\varphi \in \mathfrak{g}_1$ a Maurer-Cartan element, the set of maps
		\begin{equation}\label{twisting}
			l_{k}^{\varphi}(v_1,v_2,\dots,v_k) = \sum_{i\geq 0}\frac{1}{l!}l_{k+i}(\underbrace{\varphi,\dots,\varphi}_{i},v_1,\dots,v_k)
		\end{equation}
		defines a new $L_\infty$ structure on $\mathfrak{g}$.
		
		\paragraph{Gauge equivalence.}
		Let $\Omega_I^\bullet=\R[t,dt]$ denote the algebraic de Rham complex in one variable $t$ valued in $\R$. We consider the graded vector spaces $\mathfrak{g}[t,dt] = \mathfrak{g} \otimes_{\R} \Omega^{\bullet}_I$, which naturally inherits an $L_\infty$ structure. For $t_0 \in \mathbb{R}$, we define the evaluation map
		\begin{equation*}
			\begin{aligned}
				e_{t_0}: \mathfrak{g}[t,dt] &\to \mathfrak{g}\\
				\alpha(t) + \beta(t)dt &\mapsto \alpha(t_0)
			\end{aligned}
		\end{equation*}
		Let $\alpha_1,\alpha_2$ be two elements of $MC(\mathfrak{g})$. We say
		$$
		\boxed{\alpha_1\  \text{is gauge equivalent to}\ \alpha_2}
		$$
		if there exists $\tilde{\alpha} \in MC(\mathfrak{g}[t,dt])$ such that $e_0(\tilde{\alpha}) = \alpha_0,e_1(\tilde{\alpha}) = \alpha_1$. Explicitly, let $\tilde{\alpha} = \alpha(t) + \beta(t)dt$ be a Maurer-Cartan equation element, then
		\begin{equation*}
			(l_1 + d_t)(\alpha(t) + \beta(t)dt) + \sum_{k \geq 2}\frac{1}{k!}l_k(\alpha(t) + \beta(t)dt,\dots,\alpha(t) + \beta(t)dt) = 0
		\end{equation*}
		which is equivalent to
		\begin{eqnarray*}
			&\sum\limits_{k\geq 1}\frac{1}{k!}l_k(\alpha(t),\dots,\alpha(t)) = 0\\
			&\frac{\partial \alpha(t)}{\partial t} = d\beta(t) + \sum_{k\geq1}\frac{1}{k!}l_{k+1}(\alpha(t),\dots,\alpha(t),\beta(t)).
		\end{eqnarray*}
		$\alpha_1$ is gauge equivalent to $\alpha_2$ if and only if we can find $\alpha(t),\beta(t)$ that $\alpha(0)=\alpha_0,\alpha(1) = \alpha_1$ and satisfy the above equations. The first equation says that $\alpha(t)$ gives a family of Maurer-Cartan elements, and the second equation says that the variation of $\alpha(t)$ along $t$ is given by the infinitesimal form of gauge symmetry
		\begin{equation}\label{Gauge-trans-0}
			\delta \alpha = d\beta + \sum_{k\geq1}\frac{1}{k!}l_{k+1}(\alpha,\dots,\alpha,\beta)
		\end{equation}

		\section{Introduction to Homological perturbation theory}
		\label{Sec_HPT}
		Homological perturbation theory is concerned with transferring various kinds of algebraic structure through a homotopy equivalence. It provides us techniques for the transference of structures from one object to another.
		
		In many situation, algebras under consideration are related to physical theory. Then the homological perturbation theory has physical interpretations. Generally speaking, homological perturbation is the same as integrating out high energy modes or axillary field. Various formulas that appear in homological perturbation theory have the physical interpretation of summing over Feynman diagrams. In this section, we present an introduction to holmological perturbation theory to prepare for our applications to higher spin theory. For more about homological perturbation theory, see e.g. \cite{Crainic,Huebschmann,Huebschmann:2002}.
		\subsection{Homological perturbation lemma}

		A chain homotopy equivalence between two chain complexes $(W,d_W)$ and $(V,d_V)$ is a chain map $i:W\to V$ such that there exist a chain map $p: V \to W$, with $i\circ p$ homotopic to $\mathds{1}_V$ and $p\circ i$ homotopic to $\mathds{1}_W$.
		\begin{equation}\label{HE_data}
			\begin{aligned}
				h\curved (V,d_V)&\overset{p}{\underset{i}\rightleftarrows} (W,d_W)\anticurved h'\\
				i\circ p-\mathds{1}_V  = d_V\circ h + h\circ d_V&,\quad p\circ i - \mathds{1}_{W} = d_W \circ h' + h'\circ d_W
			\end{aligned}
		\end{equation}
		In this case,  $W$ and $V$ are called chain homotopy equivalent. If $h' = 0$, then $\mathds{1}_W = p\circ i $, $i$ is injective, $p$ is surjective and the chain complex $W$ is called a deformation retract of $V$.
		\begin{equation}\label{def_re}
			h\curved (V,d_V)\overset{p}{\underset{i}\rightleftarrows} (W,d_W)
		\end{equation}
		A deformation retract satisfying the following conditions
		\begin{equation}
			h\circ i = 0,\;p\circ h = 0,\; h\circ h = 0
		\end{equation}
		is called special deformation retract, or SDR.
		
		\begin{lem}[Homological Perturbation]\label{HPT_lemma}
			A perturbation $\delta$ of \eqref{def_re} is a map on $V$ of the same degree as $d_V$, such that $(d_V + \delta )^2= 0$. We call it small if $1-\delta h$ is invertible. Then given a perturbation $\delta$, there is a new chain homotopy equivalence
			\begin{equation*}
				h'\curved (V,d_V+\delta)\overset{p'}{\underset{i'}\rightleftarrows} (W,d'_W)
			\end{equation*}
			where the maps are given by
			\begin{align*}
				h' &= h + h(1-\delta h)^{-1}\delta h\\
				p' &= p + p(1 - \delta h)^{-1}\delta h\\
				i' & =  i +h(1 - \delta h)^{-1}\delta i\\
				d'_W & = d_W + p(1 - \delta h)^{-1}\delta i
			\end{align*}
			Moreover, when the initial data is a SDR, the perturbed data is also a SDR.
		\end{lem}
		We will focus on the SDR case in our applications of homological perturbation theory.
		
		\subsection{Transferring algebraic structure}
		\label{Sec_tree}
		Given two isomorphic vector spaces and an algebra structure on one of them, one can always define, by means of identification, an algebraic structure on the other space such that these two algebraic structures become isomorphic. Physically, however, we are not only interested in theories that are isomorphic, but also interested in theories that are related by renormalization. This usually leads to quasi-isomorphisms. Given two quasi-isomorphic chain complexes and an algebra structure on one of the chain complex, one can still define an algebraic structure on the other. But in this case, the algebraic relations are satisfied only up to homotopy. In this section we explain this philosophy in terms of the situation that an associative algebra structure will be transferred to an $A_\infty$ structure.
		
		Consider the following deformation retract
		\begin{equation*}
			h\curved (A,d_A)\overset{p}{\underset{i}\rightleftarrows} (H,d_H)
		\end{equation*}
		where $(A,d_A)$ is a differential graded algebra. That is, we have a map $\mu: A\otimes A \to A$ satisfying $\mu(\mu(a,b),c) = \mu(a,\mu(b,c))$, $\forall a,b,c \in A$, and a differential $d_A$ which is a derivation with respect to the product $\mu$: $ d_A \mu(a,b) = \mu(d_Aa,b) + (-1)^{|a|}\mu(a,d_Ab)$.
		
		We then consider transferring this algebraic structure from $A$ to $H$. First, we can define a binary operation $m_2: H^{\otimes 2} \to H$ by the formula $m_2(a,b) = p\mu(i(a),i(b))$:
		\begin{center}
			\begin{tikzpicture}[grow' = up]
			\tikzstyle{level 1}=[level distance = 2mm]
			\tikzstyle{level 2}=[sibling distance=8mm,level distance = 6mm]
			\tikzstyle{level 3}=[sibling distance=8mm,level distance = 6mm]
			\tikzstyle{level 4}=[sibling distance=8mm,level distance = 2mm]
			\coordinate
			child {edge from parent[draw=none] child {node{$m_2$}child { child { edge from parent[draw=none]}}   child { child { edge from parent[draw=none]}}}
			};
			;	\end{tikzpicture}
			\begin{tikzpicture}[grow' = up]
			\tikzstyle{level 1}=[level distance = 2mm]
			\tikzstyle{level 2}=[sibling distance=8mm,level distance = 5mm]
			\tikzstyle{level 3}=[sibling distance=8mm,level distance = 5mm]
			\tikzstyle{level 4}=[sibling distance=8mm,level distance = 2mm]
			\coordinate
			node{$p$} child {edge from parent[draw=none] child {child { child {node{$i$} edge from parent[draw=none]}}   child { child {node{$i$} edge from parent[draw=none]}}}
			};
			\node at (-0.8,0.6) {$= \;$};
			;	\end{tikzpicture}
		\end{center}
		$d_H$ can be shown to be a differential with respect to $m_2$. However, the associativity is not granted: the failure is measured by the associator as in the following picture:
		\begin{center}
			\begin{tikzpicture}[grow' = up]
			\tikzstyle{level 1}=[level distance = 2mm]
			\tikzstyle{level 2}=[sibling distance=8mm,level distance = 6mm]
			\tikzstyle{level 3}=[sibling distance=8mm,level distance = 8mm]
			\tikzstyle{level 4}=[sibling distance=8mm,level distance = 6mm]
			\coordinate
			child {edge from parent[draw=none] child {node{$m_2$} child { node{$m_2$} child child }   child { child { edge from parent[draw=none]}}}
			};
			\node at (1,0.6) {$-$};
			\end{tikzpicture}
			\begin{tikzpicture}[grow' = up]
			\tikzstyle{level 1}=[level distance = 2mm]
			\tikzstyle{level 2}=[sibling distance=8mm,level distance = 6mm]
			\tikzstyle{level 3}=[sibling distance=8mm,level distance = 8mm]
			\tikzstyle{level 4}=[sibling distance=8mm,level distance = 6mm]
			\coordinate
			child {edge from parent[draw=none] child {node{$m_2$} child    child { node{$m_2$} child child }}
			};
			\node at (1,0.6) {$=$};
			\end{tikzpicture}
			\begin{tikzpicture}[grow' = up]
			\tikzstyle{level 1}=[sibling distance=8mm,level distance = 2mm]
			\tikzstyle{level 2}=[sibling distance=8mm,level distance = 5mm]
			\tikzstyle{level 3}=[sibling distance=8mm,level distance = 8mm]
			\tikzstyle{level 4}=[sibling distance=8mm,level distance = 5mm]
			\tikzstyle{level 5}=[sibling distance=8mm,level distance = 5mm]
			\tikzstyle{level 6}=[sibling distance=8mm,level distance = 2mm]
			\coordinate
			node {$p$}child {edge from parent[draw=none] child { child {node{$i\circ p$} child{child {child{node{$i$} edge from parent[draw=none]}}  child {child{node{$i$} edge from parent[draw=none]}} }}    child { node{$i$}}}
			};
			\node at (1,0.4) {$-$};
			\end{tikzpicture}
			\begin{tikzpicture}[grow' = up]
			\tikzstyle{level 1}=[sibling distance=8mm,level distance = 2mm]
			\tikzstyle{level 2}=[sibling distance=8mm,level distance = 5mm]
			\tikzstyle{level 3}=[sibling distance=8mm,level distance = 8mm]
			\tikzstyle{level 4}=[sibling distance=8mm,level distance = 5mm]
			\tikzstyle{level 5}=[sibling distance=8mm,level distance = 5mm]
			\tikzstyle{level 6}=[sibling distance=8mm,level distance = 2mm]
			\coordinate
			node {$p$}child {edge from parent[draw=none] child { child{ node{$i$}}     child {node{$i\circ p$} child{child {child{node{$i$} edge from parent[draw=none]}}  child {child{node{$i$} edge from parent[draw=none]}} }}}
			};
			\end{tikzpicture}
		\end{center}
		We see that if $i\circ p$ equals to the identity, then $m_2$ would be associative, but this is not true in general. Since $i\circ p$ equals to the identity up to the homotopy $h$,  the obstruction of $m_2$ being associative is expected to be measured by $h$. We introduce the element $m_3: H^{\otimes 3} \to H$ by $m_3(a,b,c) = p\mu (h\mu(i(a),i(b)),i(c)) - p\mu (i(a),h \mu(i(b),i(c)))$
		\begin{center}
			\begin{tikzpicture}[grow' = up]
			\tikzstyle{level 1}=[level distance = 2mm]
			\tikzstyle{level 2}=[sibling distance=8mm,level distance = 8mm]
			\tikzstyle{level 3}=[sibling distance=5mm,level distance = 8mm]
			\tikzstyle{level 4}=[sibling distance=8mm,level distance = 8mm]
			\coordinate
			child {
				edge from parent[draw=none] child {
					node{$m_3$}
					child { child {edge from parent[draw=none]}}  child { child {edge from parent[draw=none]}} child { child {edge from parent[draw=none]}}
				}
			};
			\node at (1,0.5) {$=$}
			;	\end{tikzpicture}
			\begin{tikzpicture}[grow' = up]
			\tikzstyle{level 1}=[sibling distance=8mm,level distance = 2mm]
			\tikzstyle{level 2}=[sibling distance=8mm,level distance = 7mm]
			\tikzstyle{level 3}=[sibling distance=8mm,level distance = 7mm]
			\tikzstyle{level 4}=[sibling distance=8mm,level distance = 7mm]
			\tikzstyle{level 5}=[sibling distance=8mm,level distance = 2mm]
			\coordinate
			node {$p$}child {edge from parent[draw=none]
				child {
					child {	
						child {
							child{node{$i$} edge from parent[draw=none]}}  child {child{node{$i$} edge from parent[draw=none]}} edge from parent node[left] {$h$} }    child { node{$i$}}
				}
			};
			\node at (1,0.4) {$-$};
			\end{tikzpicture}
			\begin{tikzpicture}[grow' = up]
			\tikzstyle{level 1}=[sibling distance=8mm,level distance = 2mm]
			\tikzstyle{level 2}=[sibling distance=8mm,level distance = 7mm]
			\tikzstyle{level 3}=[sibling distance=8mm,level distance = 7mm]
			\tikzstyle{level 4}=[sibling distance=8mm,level distance = 7mm]
			\tikzstyle{level 5}=[sibling distance=8mm,level distance = 2mm]
			\coordinate
			node {$p$}child {edge from parent[draw=none]
				child { child { node{$i$} }
					child {	
						child {
							child{node{$i$} edge from parent[draw=none]}}  child {child{node{$i$} edge from parent[draw=none]}} edge from parent node[right] {$h$} }
				}
			};
			\end{tikzpicture}
		\end{center}
		Then one can check that
		\begin{equation*}
			\partial(m_3) =  m_2\circ(m_2 \otimes \mathds{1}) - m_2\circ(\mathds{1}\otimes m_2)
		\end{equation*}
		This is depicted as:
		\begin{center}
			\begin{tikzpicture}[grow' = up]
			\tikzstyle{level 1}=[level distance = 2mm]
			\tikzstyle{level 2}=[sibling distance=8mm,level distance = 6mm]
			\tikzstyle{level 3}=[sibling distance=5mm,level distance = 6mm]
			\tikzstyle{level 4}=[sibling distance=8mm,level distance = 6mm]
			\coordinate
			child {
				edge from parent[draw=none] child {
					node{$m_3$}
					child { child {edge from parent[draw=none]}}  child { child {edge from parent[draw=none]}} child { child {edge from parent[draw=none]}}
				}
			};
			\node at (1,0.7) {$\left.\right)=$};
			\node at (-1,0.7) {$\partial\left( \right.$};
			\end{tikzpicture}
			\begin{tikzpicture}[grow' = up]
			\tikzstyle{level 1}=[level distance = 2mm]
			\tikzstyle{level 2}=[sibling distance=8mm,level distance = 6mm]
			\tikzstyle{level 3}=[sibling distance=8mm,level distance = 8mm]
			\tikzstyle{level 4}=[sibling distance=8mm,level distance = 6mm]
			\coordinate
			child {edge from parent[draw=none] child {node{$m_2$} child { node{$m_2$} child child }   child { child { edge from parent[draw=none]}}}
			};
			\node at (1,0.7) {$-$};
			\end{tikzpicture}
			\begin{tikzpicture}[grow' = up]
			\tikzstyle{level 1}=[level distance = 2mm]
			\tikzstyle{level 2}=[sibling distance=8mm,level distance = 6mm]
			\tikzstyle{level 3}=[sibling distance=8mm,level distance = 8mm]
			\tikzstyle{level 4}=[sibling distance=8mm,level distance = 6mm]
			\coordinate
			child {edge from parent[draw=none] child {node{$m_2$} child    child { node{$m_2$} child child }}
			};
			\end{tikzpicture}
		\end{center}
		This means that the associator of $m_2$ vanishes up to the homotopy by $m_3$. Generalizing the previous formulas, we have the following family of maps
		\begin{equation}\label{tree-formula}
			\boxed{m_n = \sum_{T\in PBT_{n}}\pm m_T},\; \forall n \geq 2
		\end{equation}
		where the notation $PBT_n$ stands for the set of planar binary rooted trees with $n$ leaves. The operation $m_T$ is obtained by putting $i$ on the leaves, $\mu$ on the vertices, $h$ on the internal edges and $p$ on the root.
		\begin{center}
			\begin{tikzpicture}[grow' = up]
			\tikzstyle{level 1}=[level distance = 2mm]
			\tikzstyle{level 2}=[sibling distance=8mm,level distance = 8mm]
			\tikzstyle{level 3}=[sibling distance=8mm,level distance = 8mm]
			\coordinate
			child {
				edge from parent[draw=none] child {
					node{$m_n$}
					child { child {edge from parent[draw=none]}}  child { child {edge from parent[draw=none]}} child { child {edge from parent[draw=none]}} child { child {edge from parent[draw=none]}} child { child {edge from parent[draw=none]}}
				}
			};
			\node at(3,1) {$ = \sum\limits_{PBT_n}\pm$};
			\end{tikzpicture}	
			\begin{tikzpicture}[grow' = up]
			\tikzstyle{level 1}=[sibling distance=10mm,level distance = 2mm]
			\tikzstyle{level 2}=[sibling distance=10mm,level distance = 7mm]
			\tikzstyle{level 3}=[sibling distance=15mm,level distance = 7mm]
			\tikzstyle{level 4}=[sibling distance=10mm,level distance = 9mm]
			\tikzstyle{level 5}=[sibling distance=10mm,level distance = 7mm]
			\tikzstyle{level 6}=[sibling distance=10mm,level distance = 2mm]
			\coordinate
			node {$p$}child {edge from parent[draw=none]
				child {
					child {	
						child {
							node{$i$}}  child {node{$i$}} edge from parent node[left] {$h$} }    child {	
						child {
							node{$i$}}  child {	
							child {
								child{node{$i$} edge from parent[draw=none]}}  child { child{node{$i$} edge from parent[draw=none]}} edge from parent node[right] {$h$} } edge from parent node[right] {$h$} }
				}
			};
			\end{tikzpicture}
		\end{center}
		With an appropriate choice of sign, $\{m_1 = d_H,\;m_n:H^{\otimes n} \to H, n \geq 2\}$ defines an $A_\infty$ structure on $H$. We will use homological perturbation theory to show this in the next section. For more about the tree description, and its generalization to transferring $A_\infty$ structure, see e.g. \cite{Markl}.
		\subsection{Transferring via HPT}
		In this section, we explain how to use homological perturbation theory to transfer algebraic structure. For more detail, see e.g. \cite{Huebschmann}. Consider a SDR data
		\begin{equation}\label{SDR}
			h\curved (V,d_V)\overset{p}{\underset{i}\rightleftarrows} (H,d_H)
		\end{equation}
		with
		$$
		\mathds{1}_H = p\circ i,\;   i\circ p-\mathds{1}_V = d_V\circ h + h \circ d_V
		$$
		and
		$$
		h\circ i = 0,\;p\circ h = 0,\; h\circ h = 0
		$$
		
		We are interested in the case when $V$ carries further algebraic structures, e.g., associative algebra structure. We hope to transfer this structure to $H$. In this section we consider an $A_\infty$ structure on $V$, and recall that an $A_\infty$ structure is encoded in a coderivation $b$ on the reduced tensor coalgebra $\bar T^c(V[1])$. Hence one seek for a new SDR data as follows
		\begin{equation*}
			?\curved (\bar T^c(V[1]),b)\overset{?}{\underset{?}\rightleftarrows} (\bar T^c(H[1]),?)
		\end{equation*}
		Since in this section we always work with degree shifted space, we omit the suspension $s$ for a moment. The complex and corresponding maps are all assumed degree shifted. For example $V \to V[1],\;i \to s \circ i \circ s^{-1}, d \to s \circ d \circ s^{-1}, h \to s \circ h \circ s^{-1}$. The first step is to extend the SDR data \eqref{SDR} to tensor coalgebra.
		\begin{equation}\label{tensor_SDR}
			T^ch\curved (\bar T^c(V),T^cd_V)\overset{T^c p}{\underset{T^c i}\rightleftarrows} (\bar T^c(H),T^c d_H)
		\end{equation}
		where the new differential $T^cd_V, T^cd_H$ is obtained by extending the original differential by Leibniz rule
		\begin{equation*}
			T^cd = \sum_{n \geq 1} \sum_{i = 0}^{n-1} \mathds{1}^{i}\otimes d \otimes \mathds{1}^{n - i - 1}.
		\end{equation*}
		The new projection and inclusion map are defined by
		$$
		T^ci = \sum_{n\geq 1}i^{\otimes n}, T^cp = \sum_{n\geq1}p ^{\otimes n}.
		$$
		The deformation retract is defined as
		\begin{equation*}
			T^ch = \sum_{n\geq 1} \sum_{i = 0}^{n-1} \mathds{1}^{\otimes 1} \otimes h \otimes (i\circ p)^{\otimes n - i - 1}.
		\end{equation*}
		One can first check that
		\begin{equation*}
			T^cp \circ T^ci = \sum_{n\geq 1}(p\circ i)^{\otimes n} = \mathds{1}_{T^cH}.
		\end{equation*}
		We denote
		$$
		h_n = \sum_{i = 0}^{n-1} \mathds{1}^{\otimes i} \otimes h \otimes (i\circ p)^{\otimes n - i - 1}, \quad d_n = \sum_{i = 0}^{n-1} \mathds{1}^{i}\otimes d_V \otimes \mathds{1}^{n - i - 1}.
		$$
		Then
		\begin{equation*}
			\begin{aligned}
				d_n\circ h_n + h_n\circ d_n &= -\sum_{i<j}\mathds{1}^{\otimes i }\otimes h \otimes (i\circ p)^{\otimes j-i-1}\otimes d_v\circ i \circ p \otimes (i\circ p )^{\otimes n-1-j} \\
				& + \sum_{i}\mathds{1}^{\otimes i }\otimes d_V\circ h \otimes (i\circ p)^{\otimes n-i-1} +  \sum_{i<j}\mathds{1}^{\otimes i }\otimes d_V \otimes \mathds{1}^{\otimes j-i-1}\otimes h \otimes (i\circ p )^{\otimes n-1-j}\\
				& + \sum_{i<j}\mathds{1}^{\otimes i }\otimes h \otimes (i\circ p)^{\otimes j-i-1}\otimes i \circ p \circ d_V \otimes (i\circ p )^{\otimes n-1-j}\\
				& + \sum_{i}\mathds{1}^{\otimes i }\otimes h\circ d_V \otimes (i\circ p)^{\otimes n-i-1} -  \sum_{i<j}\mathds{1}^{\otimes i }\otimes d_V \otimes \mathds{1}^{\otimes j-i-1}\otimes h \otimes (i\circ p )^{\otimes n-1-j}\\
				& = \sum_{i}\mathds{1}^{\otimes i }\otimes(i\circ p - \mathds{1} )\otimes (i\circ p)^{\otimes n-i-1}\\
				& =(i\circ p)^{\otimes n} - \mathds{1}^{\otimes n}
			\end{aligned}
		\end{equation*}
		Therefore we have
		\begin{equation*}
			T^cd_V \circ T^ch + T^ch\circ T^c d_V = T^ci\circ T^c p - \mathds{1}_{T^c V}.
		\end{equation*}
		One also easily checks that
		\begin{equation*}
			T^ch\circ T^ci = 0,\; T^cp\circ T^ch = 0,\; T^ch\circ T^ch = 0.
		\end{equation*}
		Hence \eqref{tensor_SDR} is indeed a SDR. The above construction is also called the tensor trick. The $A_\infty$ structure on $V$ is given by a codifferential
		\begin{equation*}
			b = T^cd_V + b_2 + b_3+ \dots =  T^c d_V + \delta
		\end{equation*}
		satisfying $b^2 = (T^c d_V + \delta)^2 = 0$. So we regard $\delta = b_2 + b_3 + \dots$ as a perturbation and apply Lemma \ref{HPT_lemma}. It gives us a homotopy equivalence data
		\begin{equation*}
			H\curved (\bar T^c(V),b)\overset{P}{\underset{I}\rightleftarrows} (\bar T^c(H),\partial)
		\end{equation*}
		where
		\begin{align*}
			H &= T^ch + T^ch(1- \delta T^ch)^{-1}\delta T^ch\\
			P &= T^cp + T^cp(1 - \delta T^ch)^{-1}\delta T^ch\\
			I & =  T^ci + T^ci(1 - \delta T^ch)^{-1}\delta T^ci\\
			\partial & = T^cd_H + T^cp(1 - \delta T^ch)^{-1}\delta T^ci
		\end{align*}
		The codifferential $\partial$ then encodes the transferred $A_\infty$ structure on $H$.
		\begin{eg}
			We consider the case $V$ being a DGA $(V,d_V,\mu)$ as in the previous section, then $b = T^cd_V + b_2$. The transfered codifferential on $H$ is given by (Remember the implicit degree shifting in our formula)
			\begin{equation*}
				\partial = T^cd_H + T^cp(1 - b_2 T^ch)^{-1}b_2 T^ci.
			\end{equation*}
			We can write this codifferential in terms of the set of maps $m_n: H^{\otimes n} \to H$, and find
			\begin{equation*}
				m_2 = - s^{-1}\circ \text{proj}_{H[1]} \circ (T^cp(1 - b_2 T^ch)^{-1}b_2 T^ci)\circ s^{\otimes 2} = p\circ \mu\circ (i\otimes i)
			\end{equation*}
			and
			\begin{equation*}
				\begin{aligned}
					m_3 &= - s^{-1}\circ \text{proj}_{H[1]} \circ (T^cp(1 - b_2 T^ch)^{-1}b_2 T^ci)\circ s^{\otimes 3}\\
					& = \mu\circ (h\circ \mu\circ (i\otimes i) \otimes i) - \mu\circ (i\otimes h\circ \mu\circ (i\otimes i))
				\end{aligned}
			\end{equation*}
			This is precisely the formula we find in the previous section. Moreover, the whole formula for $\partial$ is equivalent to the plenary binary tree construction of $A_\infty$ structure of $H$, and the degree shifting remember the sign appearing in our formula
		\end{eg}
		
		As we will see, Homological perturbation theory will become a powerful tool for us to analyze the higher spin equation and the Vasiliev equation.
		
		\subsection{Example: Chern-Simons theory}\label{sec:CS}
		We explain the physics content of homological perturbation theory via the example of Chern-Simons theory. Let $X$ be a smooth orientable 3-dim manifold. Let $\g$ be the Lie algebra of a Lie group $G$. Chern-Simons theory on $X$ concerns with the following differential graded Lie algebra (we consider trivial principal $G$-bundle for simplicity)
		$$
		\mathcal E= \Omega^\bullet(X)\otimes \g.
		$$
		The differential is the de Rham differential $d$ on $X$, and the Lie bracket $l_2$ is induced from $\g$. The classical Chern-Simons functional (in the BRST-BV formalism) is given by \cite{Axelrod:1991vq}
		$$
		CS(\mathcal A)=\int_X \Tr {1\over 2} \mathcal A\wedge d\mathcal A+ {1\over 6}\mathcal A\wedge [\mathcal A, \mathcal A], \quad \mathcal A\in \mathcal E[1].
		$$
		The component $ \Omega^1(X)\otimes \g$ is the connection field, $ \Omega^0(X)\otimes \g$ is the ghost field for the infinitesimal gauge symmetry, and the 2-form and 3-form components are the corresponding anti-fields. The equation of motion is
		$$
		d\mathcal A+{1\over 2}[\mathcal A, \mathcal A]=0
		$$
		which is nothing but the Maurer-Cartan equation.
		
		Let us choose a metric on $X$.  Let
		$
		d^*: \mathcal E\to \mathcal E
		$
		be the adjoint of $d$, and
		$$
		\Delta:= dd^*+d^*d
		$$
		be the Laplacian.  Let $\mathbb H=\ker \Delta\subset \mathcal E$ be the subspace of harmonics. This leads to a data of special deformation retract
		\begin{equation*}
			h\curved (\mathcal E,d)\overset{p}{\underset{i}\rightleftarrows} (\mathbb H,0)
		\end{equation*}
		Here $h=-d^*{1\over \Delta}$ where ${1\over \Delta}$ is the Green's operator on forms. $h$ is precisely the propagator of Chern-Simons theory. $i$ is the natural inclusion. $p$ is the harmonic projection.
		
		Treating the Lie structure $l_2$ as a perturbation allows us to transfer the DGLA structure on $\mathcal E$ to an $L_\infty$ structure on $\mathbb H$. The tree formula \eqref{tree-formula} gives the tree level Feynman diagrams for Chern-Simons theory, and the transferred $L_\infty$-structure is a way to organize the structure of effective theory on zero modes.
		
		\section{Unfolded Higher Spin equation and Vasiliev equation}
		\subsection{Unfolded equation as Maurer-Cartan equation}
		
		The problem of constructing consistent gauge invariant theories of interacting massless fields of higher spin (HS) is one of fundamental problem in field theory. For several decades, a lot of efforts have been put to attack this problem, and it turns out that the unfolding technique is a prominent starting point towards nonlinear higher spin theories at all orders. Based on this unfolded approach \cite{Vasiliev:1988sa,Vasiliev:1990en},  the full nonlinear dynamics of higher spin fields has been constructed at the level of equations of motion in Vasiliev theory. In this section we introduce basic concept of unfolded equation. The unfolded equation is originally formulated in terms of free differential algebras, and its relation with $L_\infty$ algebra is spelled out in \cite{Vasiliev:2005zu}. Instead of using traditional language in physical literature, we explain the unfolded equation in terms of $L_\infty$ algebra which is also basis independent. The advantage is that the underlining algebraic structure becomes clear. We will find that many problems like finding the full nonlinear equations become pure algebraic, and powerful techniques like homological perturbation theory can be applied to analyze them.
		
		For an $L_\infty$ algebra $(\mathfrak{g},l_1,l_2,\dots)$, and a smooth manifold $X$, we consider differential forms on $X$ taking value in $\mathfrak{g}$, i.e.
		$$
		\mathcal{E}^{\bullet} = \Omega^{\bullet}(X)\otimes \mathfrak{g}.
		$$
		There is naturally an $L_\infty$ structure on this space as we described previously, and this can be viewed as the $L_\infty$ generalization of ordinary Chern-Simons theory discussed in Section \ref{sec:CS}.  Then what physicists called {\textbf{unfolded equation}} is just the Maurer-Cartan equation associated to this $L_\infty$ algebra:
		\begin{equation}
			\mathcal{F} = d_x\Psi + \sum_{k\geq 1} \frac{1}{k!}l_{k}(\Psi^{\otimes k}) = 0,\quad \Psi \in \mathcal{E}^1
		\end{equation}
		Here $d_x$ is the de-Rham differential. We check that the field equation $\mathcal{F}$ satisfies the consistency condition $d_x^2 = 0$
		\begin{equation*}
			\begin{aligned}
				d_x^2\Psi&= d_x(-\sum_{m\geq 1} \frac{1}{m!}l_{m}(\Psi^{\otimes m}))\\
				&=  - \sum_{i = 1}^{m}\sum_{m\geq 1}(-1)^{2-m + i - 1}\frac{1}{m!}l_{m}(\underbrace{\Psi,\dots,d_x\Psi}_{i},\dots,\Psi)\\
				& =\sum_{m\geq 0}\sum_{k\geq 1} (-1)^{m+1}\frac{1}{m!}\frac{1}{k!}l_{m+1}(l_{k}(\Psi^{\otimes k}),\Psi^{\otimes m })\\
				& =\sum_{n\geq k}(-1)^{n+1}\frac{1}{n!}\sum_{k\geq 1}^n (-1)^{k}\binom{n}{k}l_{n-k+1}(l_{k}(\Psi^{\otimes k}),\Psi^{\otimes n-k })\\
				& = 0
			\end{aligned}
		\end{equation*}
		We used $L_\infty$ relation in the last line. This shows that the $L_\infty$ structure automatically implies the consistent condition $d_x^2 = 0$. The gauge transformations can be analyzed similarly. We use the infinitesimal form of gauge symmetry of Maurer-Cartan equation introduced earlier \eqref{Gauge-trans-0},
		\begin{equation}
			\delta_\Lambda \Psi = d_x\Lambda + \sum_{k\geq 0}\frac{1}{k!}l_{k+1}(\Psi^{\otimes k},\Lambda),\; \text{ for }\Lambda \in \mathcal{E}^{0}
		\end{equation}
		For $\Psi$ a MC element, we know from \eqref{twisting} that the twisted maps
		\begin{equation}
			\begin{aligned}
				l_1^{\Psi}(X) &= d_xX + \sum_{k\geq 0}\frac{1}{k!}l_{k+1}(\Psi^{\otimes k},X)\\
				l_{i}^{\Psi}(X_1,\dots,X_i) &= \sum_{k\geq 0}\frac{1}{k!}l_{k+i}(\Psi^{\otimes k},X_1,\dots,X_i),\; \text{ for }i>1
			\end{aligned}
		\end{equation}
		define a new $L_\infty$ algebra structure on $\mathcal{E}^{\bullet}$. Observe $\delta_\Lambda\Psi = l_1^{\Psi}(\Lambda)$. Using these properties, we can check that the field equation is invariant under the gauge transformation
		\begin{equation*}
			\begin{aligned}
				\delta_\Lambda\mathcal{F} &= l_1^{\Psi}(\delta_\Lambda\Psi) = l_1^{\Psi}(l_1^{\Psi}(\Lambda)) = 0.
			\end{aligned}
		\end{equation*}
		
		\begin{eg}
			If $(\mathfrak{g},[-,-])$ is an ordinary Lie algebra, then $\mathcal{E}^{\bullet} =\Omega^{\bullet}(X)\otimes  \mathfrak{g} $ is the space of Lie algebra valued form on $X$. The unfolded equation becomes the zero curvature condition for the one form $A \in \Omega^1(X)\otimes \mathfrak{g}$
			\begin{equation*}
				d_xA + \frac{1}{2}[A,A] = 0.
			\end{equation*}
			The gauge transformation also takes the familiar form
			\begin{equation*}
				\delta_\xi A = d_x\xi + [A,\xi].
			\end{equation*}
			The unfolded equation in this case is the equation of motion of Chern-Simons theory.
		\end{eg}
		\paragraph{Unfolded equation in physical literature}
		For an $L_\infty$ algebra $\mathfrak{g}$ with a basis $\{e_{\mathcal{A}}\}$, one can write a general element of $\mathcal{E}$ as $\Psi = \sum_{\mathcal{A}} e_{\mathcal{A}} W^{\mathcal{A}}(x)$, where $\mathcal{W}^{\mathcal{A}}$ is a set of differential forms. Then the above unfolded equation takes the following form
		\begin{equation}
			d\mathcal{W}^{\mathcal{A}} = \mathcal{F}^{\mathcal{A}}(\mathcal{W}) = \sum_{n}\sum_{|\mathcal{B}_1|+\dots |\mathcal{B}_n| = 1 + |\mathcal{A}|} f^{\mathcal{A}}_{\mathcal{B}_1,\dots \mathcal{B}_n}\mathcal{W}^{\mathcal{B}_1}\wedge\dots \mathcal{W}^{\mathcal{B}_n}
		\end{equation}
		where $f^{\mathcal{A}}_{\mathcal{B}_1,\dots \mathcal{B}_n}$ are structure constant determined by the $e_{\mathcal{A}}$ component of $l_n(e_{\mathcal{B}_1},\dots,e_{\mathcal{B}_n})$. This is the familiar form of unfolded equation in physical literature. The consistency conditions
		\begin{equation}\label{str_jacobi}
			d^{2}\mathcal{W}^{\mathcal{A}} = \mathcal{F}^{\mathcal{B}}\wedge \frac{\delta\mathcal{F}^{\mathcal{A}}}{\delta W^{\mathcal{B}} }= 0
		\end{equation}
		is guaranteed by the $L_\infty$ relation as we explained previously. We have gauge symmetry
		\begin{equation}
			\delta W^{\mathcal{A}} = d\epsilon^{\mathcal{A}} - \epsilon^{\mathcal{B}}\partial_{\mathcal{B}}\mathcal{F}^{\mathcal{A}}
		\end{equation}
		where $\epsilon^{\mathcal{A}}$ is a differential form of degree $|\mathcal{A}| - 1$.
		
		\subsection{Unfolded linearized higher spin equation}
		The linearized equation of higher spin field has been known for a long time. In this paper we only consider 4d bosonic higher spin theory, and in this section we introduce the unfolded equation for linearized higher spin field in $AdS$ background. Here we briefly review the relevant data, and we refer to \cite{Vasiliev:1999ba,Didenko:2014dwa} for detailed discussion.
		
		\paragraph{Weyl algebra}
		Given a vector space $V$, $\dim(V) = 2n$, we denote its ring of formal functions by
		\begin{equation}
			\widehat{\mathcal{O}}(V) := \widehat{S}(V^*) = \prod_{k\geq0}S^k(V^*)
		\end{equation}
		which has a natural commutative product. The Weyl algebra can be regarded as a quantization of this commutative algebra. Choose a basis $e_i$ for $V$ and let $y^i\in V^*$ be the dual basis. Then $\widehat{\mathcal{O}}(V)$ can be regarded as the space of functions $f(y_1,\dots,y_{2n})$. Let $\hbar$ be a formal parameter, and we consider the space $\widehat{\mathcal{O}}(V)[[\hbar]]$. To define the Weyl algebra, assume $V$ is endowed with a linear symplectic pairing $\omega$ of the form
		\begin{equation*}
			\omega = \frac{1}{2}\omega_{ij}dy^i\wedge dy^j.
		\end{equation*}
		This symplectic form induces a bilinear Poisson bracket denoted by $\{-,-\}$. We define the  Moyal star product $\star$ on $\widehat{\mathcal{O}}(V)[[\hbar]]$:
		\begin{equation}
			(f\star g)(y,\hbar) = \exp(\frac{\hbar}{2}\omega^{ij}\frac{\partial}{\partial y_1^i}\frac{\partial}{\partial y_2^j})f(y_1,\hbar)g(y_2,\hbar)|_{y_1=y_2 = y}
		\end{equation}
		where $(\omega^{ij})$ is the inverse matrix of $(\omega_{ij})$. This definition does not depend on the choice of basis. It is straightforward to check that the Moyal product $\star$ is associative and its first order non-commutativity is measured by the Poisson bracket
		\begin{equation*}
			\lim_{\hbar \to 0} \frac{1}{\hbar}(f\star g - g\star f) = \{f,g\} ,\quad f,g \in \widehat{\mathcal{O}}(V)[[\hbar]].
		\end{equation*}
		We define the Weyl algebra $A_n$ to be the associative algebra $(\widehat{\mathcal{O}}(V)[[\hbar]],\star)$. Alternately, the Weyl algebra $A_n$ can be defined by the tensor algebra $T(V^*)$ module the relations
		\begin{equation}
			y^{i}y^{j} - y^{j}y^{i} = \hbar \omega^{ij}.
		\end{equation}
		The equivalence between the two definition is established by choosing a symmetric ordering of the generator.
		
		\paragraph{Higher spin algebra $\mathfrak{hs}$.} We introduce the higher-spin algebra $\mathfrak{hs}$, which is the even part of the Weyl algebra $A_2$ with four generators.
		
		We use the notations as in physical literature. Denote the basis of $V^*$ as $y^{\alpha},\bar{y}^{\dot{\alpha}},(\alpha,\dot{\alpha} = 1,2)$. Our convention for the symplectic form is
		$$
		\omega^{\alpha\beta} = 2\epsilon^{\alpha\beta},\omega^{\dot{\alpha}\dot{\beta}} = 2\epsilon^{\dot{\alpha}\dot{\beta}}, \quad \text{where}\quad \epsilon^{12}=-\epsilon^{21}=1.
		$$
		So we have the following relations:
		\begin{equation}
	\begin{aligned}
	&y^\alpha \star y^\beta = y^\alpha  y^\beta + \hbar \epsilon^{\alpha\beta},\; [y^\alpha,y^\beta] = 2\hbar\epsilon^{\alpha\beta}\\
	& \bar{y}^{\dot{\alpha}}\star \bar{y}^{\dot{\beta}} =  \bar{y}^{\dot{\alpha}} \bar{y}^{\dot{\beta}} +\hbar\epsilon^{\dot{\alpha}\dot{\beta}},\;[\bar{y}^{\dot{\alpha}},\bar{y}^{\dot{\beta}}] =  2\hbar\epsilon^{\dot{\alpha}\dot{\beta}}
	\end{aligned}
		\end{equation}
		Spinor indices are raised and lowered as follows
		\begin{equation*}
		u^\alpha = \epsilon^{\alpha \beta}u_\beta,\quad u_\alpha = u^{\beta}\epsilon_{\beta\alpha}, \quad \epsilon_{12} = -\epsilon_{21} = 1.
		\end{equation*}
		In this paper we use $Y^A$ to collectively denote $y^{\alpha},\bar{y}^{\dot{\alpha}}$, and use the notation
		\begin{equation*}
			\epsilon^{AB} = \begin{pmatrix}
				\epsilon^{\alpha\beta}&0\\0&\epsilon^{\dot{\alpha}\dot{\beta}}
			\end{pmatrix}.
		\end{equation*}
		So we write $[Y^A,Y^B] = 2\hbar\epsilon^{AB}$. The Moyal product in this notation becomes
		\begin{equation*}
			(f\star g)(Y) = \exp(\hbar\epsilon^{AB}\frac{\partial}{\partial Y_1^A}\frac{\partial}{\partial Y_2^B}) (f(Y_1)g(Y_2))|_{Y_i=Y}.
		\end{equation*}
		There is also an integral formula for the Moyal product that we will use
		\begin{equation}
			(f\star g)(Y) = \int \frac{d^{4}U}{(2\pi)^2}\frac{d^{4}V}{(2\pi)^2} f(Y+U)g(Y+V)\exp^{-U_AV^A/\hbar}.
		\end{equation}
		By definition elements of HS algebra $\mathfrak{hs}$ are identified with even elements $f(-Y) = f(Y)$.
		\paragraph{Higher spin equation on AdS background}
		The quadratic monomial in $\mathfrak{hs}$ can be shown to form the Anti-de Sitter algebra $\mathfrak{so}(3,2)$. We denote
		\begin{equation}
			L_{\alpha\beta} = \frac{1}{2\hbar}y_{\alpha}y_{\beta},\quad\bar{L}_{\dot{\alpha}\dot{\beta}} = \frac{1 }{2\hbar}\bar{y}_{\dot{\alpha}}\bar{y}_{\dot{\beta}}, \quad P_{\alpha\dot{\beta}} = \frac{1}{2}y_\alpha \bar{y}_{\dot{\beta}}
		\end{equation}
	They satisfy the $\mathfrak{so}(3,2)$ commutation relation
	\begin{equation*}
	\begin{aligned}
	[L_{\alpha_1\alpha_2},L_{\beta_1\beta_2}] &= \epsilon_{\alpha_1\beta_1}L_{\alpha_2\beta_2} + \epsilon_{\alpha_1\beta_2}L_{\alpha_2\beta_1} + \epsilon_{\alpha_2\beta_1}L_{\alpha_1\beta_2} + \epsilon_{\alpha_2\beta_2}L_{\alpha_1\beta_1}\\
	[L_{\alpha_1\alpha_2},P_{\beta\dot{\beta}}]& = \epsilon_{\alpha_1\beta}P_{\alpha_2\dot{\beta}}+ \epsilon_{\alpha_2\beta}P_{\alpha_1\dot{\beta}}\\
	[P_{\alpha\dot{\alpha}}, P_{\beta\dot{\beta}}]& = \hbar^2(\epsilon_{\alpha\beta}\bar{L}_{\dot{\alpha}\dot{\beta}}+\bar{\epsilon}_{\dot{\alpha}\dot{\beta}}L_{\alpha\beta})
	\end{aligned}
	\end{equation*}
	From above we see that $\hbar^{-1}$ is identified with $AdS$ radius. Then the background frame field $h^{\alpha\dot{\alpha}}$ and spin connection $\varpi^{\alpha\dot{\alpha}}$ can be packed into a $\mathfrak{hs}$ valued $1$-form
		\begin{equation}
			\Omega = \frac{1}{2}\varpi^{\alpha\beta}L_{\alpha\beta} + h^{\alpha\dot{\alpha}}P_{\alpha\dot{\alpha}} + \frac{1}{2}\bar{\varpi}^{\dot{\alpha}\dot{\beta}}\bar{L}_{\dot{\alpha}\dot{\beta}}
		\end{equation}
		The zero curvature condition for the above connection $d\Omega + \Omega\star\Omega = 0$ is just the defining equation for $AdS_4$ vacuum.
		
		The field content of unfolded higher spin equation can be packed into a flat connection $\Omega$, a $\mathfrak{hs}$ valued 1-form $w$ and a $\mathfrak{hs}$ valued 0-form $C$. The unfolded equations for linearized higher spin field on $AdS_4$ background have the following form:
		\begin{framed}
			\begin{equation}\label{linearized_HS}
				\begin{aligned}
					&dw + \Omega\star w + w\star \Omega + \mathcal{V}(\Omega,\Omega,C)=0 \\
					&dC + \Omega\star C - C\star\pi(\Omega) = 0
				\end{aligned}
			\end{equation}
		\end{framed}
	\noindent	where $\pi$ is an automorphism of the Weyl algebra:
		$$
		\pi(f)(y,\bar{y}) := f(-y,\bar{y}).
		$$
		The vertex $\mathcal{V}(\Omega,\Omega,C)$ is given by
		\begin{equation}\label{hs_gluing}
			\mathcal{V}(\Omega,\Omega,C) = -h^{\gamma\dot{\alpha}}\wedge h_{\gamma}^{\dot{\beta}}\frac{\partial^2}{\partial \bar{y}^{\dot{\alpha}} \partial\bar{y}^{\dot{\beta}}}C(0,\bar{y}) - h^{\alpha\dot{\gamma}}\wedge h^{\beta}_{\dot{\gamma}}\frac{\partial^2}{\partial y^{\alpha} \partial y^{\beta}}C(y,0).
		\end{equation}
		
		The automorphism $\pi$ is an important ingredient of HS theories. It can be realized as an inner automorphism through the star product. Define the following Klein operators
		\begin{equation}
			\varkappa_y = 2\pi\delta^2(y),\quad \bar{\varkappa}_y = 2\pi\delta^2(\bar{y}).
		\end{equation}
		Then one can check that
		\begin{equation}
			\varkappa_y \star \varkappa_y = 1,\quad \varkappa_y\star f = \pi(f)\star\varkappa_y.
		\end{equation}
		Strictly speaking the Klein operator does not belong to the Weyl algebra. Nevertheless, the star product between a polynomial function and the Klein operator is defined using the integral formula. We have written down the free equation of motion for higher spin field. The next step is to turn on the interaction, or to find possible nonlinear completion of the equations. We will analyze the algebraic nature of this problem in the next section.
		
		\subsection{Nonlinear higher spin equations}
		So far we have introduced the unfolded form of free equation for higher spin field and they have interesting structure related to Weyl algebra. If we believe in the existence of the consistent nonlinear higher spin theory then these equations must be the linearizion of the following nonlinear unfolded equation
		\begin{framed}
			\begin{equation}
				\begin{aligned}
					0 &= d w + \mathcal{V}(w,w) + \mathcal{V}(w,w,C) + \mathcal{V}(w,w,C,C) \dots\\
					0 &= d C + \mathcal{V}(w,C) + \mathcal{V}(w,C,C) + \mathcal{V}(w,C,C,C) \dots
				\end{aligned}
			\end{equation}
		\end{framed}
		\noindent where the vertices $\mathcal{V}$ correspond to certain interaction and need to obey the integrability condition of the unfolded equation. Indeed, if we substitute $w \to \Omega + gw,\;C \to gC$ and pick up the terms of the first order in $g$, we can match the form of the linearized equation, and the known linear equation tells us that
		\begin{equation}
			\mathcal{V}(w,w) = w\star w,\quad \mathcal{V}(w,C) = w\star C - C\star \pi(w)
		\end{equation}
		As a trivial example, we have the following equation that satisfies integrability condition
		\begin{equation}
			\begin{aligned}\label{hs_trivial}
				&0 = d w + \mathcal{V}(w,w) \\
				&0 =  d C + \mathcal{V}(w,C)
			\end{aligned}
		\end{equation}
		However, this cannot be the physical theory since it does not give us the gluing term $\mathcal{V}(\Omega,\Omega,w)$ derived from free higher spin equation. Therefore, we look for nonlinear deformation of the trivial equation \eqref{hs_trivial}, subject to the condition \eqref{hs_gluing}. As we explained, the unfolded equation is nothing but the Maurer-Cartan equation of certain $L_\infty$ algebra. Then the deformation of an unfolded equation is mathematically the deformation of certain $L_\infty$ structure. Let us first write down the $L_\infty$ structure of the unfolded HS equation.
		
		If we regard the hgher spin algebra $\mathfrak{hs}$ as a Lie algebra, then the corresponding MC equation is the zero curvature equation for the HS connection: $dw + w\star w = 0$, where we miss the zero form $C$. To incorporate the zero form, we introduce a formal symbol $\epsilon$ of degree $1$ and square to zero $\epsilon\epsilon = 0$. We consider the space
		$$
		\mathfrak{hs}[\epsilon] = \{a+b\epsilon| a,b\in\mathfrak{hs}\}
		$$
		and extend our Moyal product $\star$ to an algebraic structure on $\mathfrak{hs}[\epsilon]$  by defining
		\begin{equation}
			\epsilon\star\epsilon = 0,\quad f\star \epsilon = f\epsilon,\quad \epsilon \star f = \pi(f)\epsilon.
		\end{equation}
		It is easy to check that this defines an associative algebra structure on $\mathfrak{hs}[\epsilon]$.
		\begin{remark}
			Note that the Klein operator and the formal symbol $\epsilon$ both generate the $\pi$ automorphism. We emphasis that they are not the same. $\epsilon$ have degree $1$ and square to $0$, which is only a formal symbol and is added to the algebra by hand so that we can compactly organize the unfolded HS equation into a single Maurer-Cartan equation.
		\end{remark}
		
		By anti-symmetrization, we naturally have a Lie bracket on $\mathfrak{hs}[\epsilon]$ given by
		\begin{equation*}
			[f_1, f_2] = [f_1,f_2],\;[f_1,f_2\epsilon] = (f_1\star f_2 - f_2\star\pi(f_1))\epsilon ,\; [f_1\epsilon,f_2\epsilon] = 0,\quad f_1,f_2\in\mathfrak{hs}
		\end{equation*}
		The Maurer-Cartan equation for the DGLA $\mathcal{E}^{\bullet} = \Omega^{\bullet}(X)\otimes \mathfrak{hs}[\epsilon]$ is
		\begin{equation*}
			d_x\Psi + \frac{1}{2}[\Psi,\Psi] = 0
		\end{equation*}
		where $\Psi \in\mathcal{E}^{1} = (\Omega^{1}(X)\otimes\mathfrak{hs} )\oplus (\Omega^{0}(X)\otimes\mathfrak{hs}\epsilon)$. If we write explicitly $\Psi = w + C \epsilon$ in components, the Maurer-Cartan equation becomes two equations
		\begin{equation*}
			\begin{aligned}
				&0 = d w + w\star w\\
				&0 =  d C + w\star C - C\star \pi(w)
			\end{aligned}
		\end{equation*}
		This tells us that the Maurer-Cartan equation for the DGLA $\mathcal{E}^{\bullet} = \Omega^{\bullet}(X)\otimes \mathfrak{hs}[\epsilon]$ is the unphysical and ``trivial" one \eqref{hs_trivial}. Then to find the right full nonlinear HS equation, we face the following algebraic problem
		
		\begin{framed}
			\noindent \textbf{Problem}: Deform the Lie algebra $\mathfrak{hs}[\epsilon]$ into an $L_\infty$ algebra, so that $l_3$ reproduce the term \eqref{hs_gluing} and the higher operations give us a set of consistent equations.
		\end{framed}

		There are two approaches to this deformation problem. One is to construct the vertex by hands and check the consistent equation order by order. Following this line, the cubic vertex is constructed \cite{Vasiliev:1988sa}. However the higher order vertex is hard to construct. The other approach is to first introduce auxiliary variables $Z^A$, and write down equations whose consistent condition is easily checked and reproduce the linearized equation upon reducing the auxiliary variables $Z^A$. This approach is the Vasiliev higher spin theory and we will introduce it in the next.

		\subsection{Vasiliev equation}
		The key elements for the construction of Vasiliev equations consist of the doubling of variable from $Y$ to $Y,Z$ and a nontrivial star-product that mixes $Y$ and $Z$. Denote $V$ the vector space whose dual $V^*$ has basis $Y^A$. Consider a copy $V_Z$ of $V$. Denote a basis of $V_Z^*$ by $Z^A = (z^\alpha,\bar{z}^{\dot{\alpha}})$. We consider the space of formal function $\widehat{\mathcal{O}}(V\oplus V_Z)$ on the doubled space $V\oplus V_Z$. The vector space $V$ is equipped with a symplectic form $\omega^{AB} = 2\epsilon^{AB}$ as before, and as a copy, $V_Z$ have the same symplectic structure. Hence $V\oplus V_Z$ naturally has a symplectic structure. However, instead of the standard Moyal product associated to this symplectic form, we give $\widehat{\mathcal{O}}(V\oplus V_Z)$ the following star product	
		\begin{equation}\label{new-star}
			(f\star g)(Y,Z) = \exp(\hbar\epsilon^{AB}(\frac{\partial}{\partial Y_1^A} + \frac{\partial}{\partial Z_1^A})(\frac{\partial}{\partial Y_2^A} - \frac{\partial}{\partial Z_2^A})) (f(Y_1,Z_1)g(Y_2,Z_2))|_{Y_i = Y,Z_i = Z}
		\end{equation}
		which has an integral expression
		\begin{equation}
			(f\star g)(Y,Z) = \int \frac{d^4 U}{(2\pi)^2} \frac{d^4V}{(2\pi)^2} f(Y+U,Z+U)g(Y+V,Z-V) e^{-U_A V^A/\hbar}.
		\end{equation}
		We have
		\begin{equation*}
			[Z^A,Z^B]_\star = 2\hbar\epsilon^{AB},\quad [Z^A,Y^B]_\star = 0.
		\end{equation*}
		
		The field content of the Vasiliev equation is given by the following set of fields all taking values in the extended algebra $(\widehat{\mathcal{O}}(V\oplus V_Z),\star)$:
		\begin{enumerate}
			\item Gauge connection $W = W_{\mu }(x|Y,Z)dx^{\mu}$, whose value at $Z=0$ gives the connection one form of the higher-spin system $w=w_\mu(x|Y) dx^{\mu}$. The bosonic projection implies $W(x|Y,Z)=W(x|-Y,-Z)$.
			\item Zero-form $B=B(x|Y,Z)$, whose value at $Z=0$ gives the zero-form of the higher-spin system $C=C(x|Y)$. The bosonic projection implies $B(x|Y,Z)=B(x|-Y,-Z)$.
			\item Auxiliary field $\mathcal{A}=\mathcal{A}_\alpha(x|Y,Z) dz^{\alpha} + \mathcal{A}_{\dot{\alpha}}(x|Y,Z)d\bar{z}^{\dot{\alpha}}$, viewed as a one-form in the auxiliary $Z$-space. We require that
			\begin{equation*}
				\mathcal{A}_\alpha(x|-Y,-Z) = -\mathcal{A}_\alpha(x|Y,Z),\quad \; \mathcal{A}_{\dot{\alpha}}(x|-Y,-Z) = - \mathcal{A}_{\dot{\alpha}}(x|Y,Z).
			\end{equation*}
		\end{enumerate}
		We can combine $W$ and $\mathcal{A}$ into a single one form $\mathcal{W} = W + \mathcal{A}$. The Vasiliev equation takes the following form:
		\begin{framed}
			\begin{equation}
				\begin{aligned}
					&d\mathcal{W} + \mathcal{W}\star \mathcal{W} = dz^2 V(B\star\varkappa)  + d\bar{z}^2\bar{V}(B\star\bar{\varkappa})\\
					&dB + 	\mathcal{W}\star B - B\star \pi(\mathcal{W})=0
				\end{aligned}
			\end{equation}
		\end{framed}
		\noindent Here $d = d_x + d_Z$. The automorphism $\pi$ is defined by
		\begin{equation*}
			\pi(y,z,dz) = (-y,-z,-dz),\;\pi(\bar{y},\bar{z},d\bar{z}) = (\bar{y},\bar{z},d\bar{z})
		\end{equation*}
		We define the Klein operator $\varkappa = \varkappa_y\star\varkappa_z$, which can be calculated to be
		\begin{equation}
			\varkappa = \int d^2ud^2v \delta^2(y + u)\delta^2(z - v)e^{- u_\alpha v^\alpha/\hbar} = e^{-z_\alpha y^\alpha/\hbar}.
		\end{equation}
		Similarly, $\bar{\varkappa} = e^{-\bar{z}_{\dot{\alpha}}\bar{y}^{\dot{\alpha}}/\hbar}$. Note that $\varkappa\star f \star \varkappa = \pi(f)$. $V,\bar{V}$ appeared in the equations are arbitrary star product function $V(X) = c_0 + c_1X+ c_2X\star X\dots$. In this paper we will focus on the case that $V$ only has linear term $V(X) = c_1X,\bar{V}(X) = \bar{c}_1X$.
		
		The consistent condition of Vasiliev equations can be checked by direct computation. In the next subsection, we will construct the $L_\infty$ structure of Vasiliev equation. The consistent condition will follow automatically.

		\subsection{$L_\infty$ structure of Vasiliev equation}
		We focus on the case that $V(X)$ is a linear function in Vasiliev equation, then the associated $L_\infty$ structure is in fact a DGLA, with vanishing $l_k$ for $k\geq 3$. First we consider the vector space of formal function of $Y,Z$ and differential form on $Z$ space
		$$
		\widehat{\mathcal{O}}(V\oplus V_Z)\otimes \wedge^{\bullet}(V_Z^*).
		$$
		A general element of this space can be written as
		\begin{equation*}
			\alpha = \alpha_{A_1\dots A_q}(Y,Z)dZ^{A_1}\wedge \cdots\wedge dZ^{A_q}.
		\end{equation*}
		The product structure on this space is induced from the wedge product on forms and star product \eqref{new-star} on formal function of $(Y,Z)$
		\begin{equation}
			\alpha\star \beta = \alpha_{A_1\dots A_q}(Y,Z)\star \beta_{B_1\dots B_p}(Y,Z) (dZ^{A_1}\wedge \cdots\wedge dZ^{A_q})\wedge (dZ^{B_1}\wedge \cdots\wedge dZ^{B_p})
		\end{equation}
		There is a natural grading on this space by requiring $\deg dZ^A = 1$. As before, we need to add a twisting element $\epsilon$ of degree $1$ to incorporate the zero form into our equation. Therefore we consider the space
		$$
		\widehat{\mathcal{O}}(V\oplus V_Z)\otimes \wedge^{\bullet}(V_Z^*)[\epsilon]
		$$
		and extend our product structure by
		\begin{equation}
			\epsilon\star\epsilon = 0,\; \alpha\star\epsilon = \alpha\epsilon,\; \epsilon \star\alpha = (-1)^{\deg \alpha}\pi(\alpha) \epsilon.
		\end{equation}
		Here we extend the automorphism $\pi$ on differential form by $\pi(dz,d\bar{z}) = (-dz,d\bar{z})$. This defines an associative product.
		
\begin{defn}		We define the algebra $\mathcal{A}^{\bullet}$ to be the even part of $\widehat{\mathcal{O}}(V\oplus V_Z)\otimes \wedge^{\bullet}(V_Z^*)[\epsilon]$ under the following $\mathbb{Z}_2$ action:
		\begin{equation}
			Y,Z,dZ \to -Y,-Z,-dZ.
		\end{equation}
\end{defn}		
		We consider the differential $d_Z$ on $\mathcal{A}^{\bullet}$ which extends the de Rham differential on $Z$-variables such that
		\begin{equation*}
			d_Z(\alpha \epsilon) = d_Z(\alpha)\epsilon.
		\end{equation*}
		The graded Leibniz rules are satisfied:
		\begin{equation*}
			\begin{aligned}
				d_Z(\alpha\star\beta) &= d_Z(\alpha)\star \beta + (-1)^{\deg\alpha}\alpha\star d_Z\beta\\
				d_Z((\alpha\epsilon)\star\beta) &= (-1)^{\deg\beta}d_Z(\alpha\star\pi(\beta)\epsilon) \\
				&= (-1)^{\deg\beta}d_Z(\alpha)\star\pi(\beta)\epsilon + (-1)^{\deg\beta + \deg\alpha}\alpha\star\pi(d_Z\beta)\epsilon\\
				&=d_Z(\alpha\epsilon)\star\beta + (-1)^{\deg\alpha +1}(\alpha\epsilon)\star d_Z\beta
			\end{aligned}
		\end{equation*}
		Thus we have a DGA $(\mathcal{A}^\bullet,d_Z,\star)$, and we denote the associated DGLA by $(\mathfrak{g}^{\bullet},d_Z,[\;,\;])$. The Maurer-Cartan equation of the DGLA $\Omega^{\bullet}(X)\otimes \mathfrak{g}^\bullet$ gives
		\begin{equation*}
			(d_x + d_Z)\Psi + \frac{1}{2}[\Psi,\Psi] = 0.
		\end{equation*}
		We can expand the field $\Psi = \mathcal{W} + B\epsilon$, and find that the above equation is the Vasiliev equation with $V = \bar{V} = 0$
		\begin{equation*}
			\begin{aligned}
				&d \mathcal{W} +  \mathcal{W}\star  \mathcal{W} = 0\\
				&dB + 	 \mathcal{W}\star B - B\star \pi( \mathcal{W})=0
			\end{aligned}
		\end{equation*}
		It is now natural to expect that upon reduction of $Z$, this equation reduces to the undeformed higher spin equation \eqref{hs_trivial}. This is indeed the case and we will show this later.
		
		To get the full nonlinear unfolded higher spin equation, we need to add structures to the DGLA $(\mathfrak{g}^{\bullet},d_Z,[\;,\;])$ so that we have nonzero $V,\bar{V}$ in our equation. Consider perturbing the differential by a $d_\epsilon$ which is defined as follows:
		\begin{equation}
			d_\epsilon(Y,Z,dZ) = 0,\;d_\epsilon(\epsilon) = c_1 dz^2 \varkappa + \bar{c}_1 d\bar{z}^2\bar{\varkappa}
		\end{equation}
		Denote $\mathcal{K} = c_1 dz^2 \varkappa + \bar{c}_1 d\bar{z}^2\bar{\varkappa}$. In general for differential form $\alpha,\beta$ we have 
		\begin{equation}
			d_{\epsilon}(\alpha + \beta\epsilon )= (-1)^{\deg \beta}\beta \star \mathcal{K}
		\end{equation}
		The graded Leibniz rule can be checked:
		\begin{equation*}
			\begin{aligned}
				d_\epsilon(\alpha\star\beta) &= 0 = d_\epsilon(\alpha) \star \beta + (-1)^{\deg \alpha}\alpha \star d_\epsilon(\beta)\\
				d_\epsilon(\alpha\star(\beta\epsilon)) & = (-1)^{\deg\alpha + \deg\beta}\alpha\star\beta \star\mathcal{K} = d_\epsilon(\alpha) \star \beta\epsilon + (-1)^{\deg \alpha}\alpha \star d_\epsilon(\beta\epsilon)\\
				d_\epsilon((\alpha\epsilon)\star\beta) & = (-1)^{\deg \alpha} \alpha\star\pi(\beta)\star\mathcal{K} = (-1)^{\deg \alpha} \alpha\star\mathcal{K}\star\beta \\
				&= d_\epsilon(\alpha\epsilon) \star \beta + (-1)^{\deg \alpha + 1}(\alpha\epsilon) \star d_\epsilon(\beta)
			\end{aligned}
		\end{equation*}
		We find that
		$$
		d_\epsilon^2 = 0,\quad d_\epsilon d_Z + d_Z d_\epsilon = 0, \quad (d_Z + d_\epsilon)^2 = 0.
		$$
		Therefore we obtain a new DGA $(\mathcal{A}^{\bullet},d_Z + d_\epsilon,\star)$ and a new DGLA $(\mathfrak{g}^{\bullet},d_Z + d_\epsilon,[\;,\;])$. The Maurer-Cartan equation associated to the new DGLA  is
		\begin{equation}
			\begin{aligned}
				&d \mathcal{W} +  \mathcal{W}\star  \mathcal{W}= (c_1dz^2 B\star\varkappa  + \bar{c}_1d\bar{z}^2B\star\bar{\varkappa})\\
				&dB + 	 \mathcal{W}\star B - B\star \pi( \mathcal{W})=0
			\end{aligned}
		\end{equation}
		which corresponds to $V(X) = c_1X,\bar{V}(X) = \bar{c}_1X$ as we want. Much information about the Vasiliev equation is encoded in the algebraic structure of the DGLA $(\mathfrak{g}^{\bullet},d_Z + d_\epsilon,[-,-])$. In particular, the consistent condition of Vasiliev equations follows automatically.
		\begin{remark}
			Although we will only consider the case for $V(X) = c_1X,\bar{V}(X) = \bar{c}_1X$, we can easily incorporate general cases by adding higher product for $\mathcal{A}^{\bullet}$. For example, we can define the following $\mu_3$:
			\begin{equation*}
				\begin{aligned}
					\mu_3(f_1\epsilon,f_2\epsilon,f_3\epsilon) & = c_3 dz^2 (f_{1}\star \varkappa)\star(f_{2}\star \varkappa)\star(f_{3}\star \varkappa)\\
					& + \bar{c}_3d\bar{z}^2(f_{1}\star \bar{\varkappa})\star(f_{2}\star \bar{\varkappa})\star(f_{3}\star \bar{\varkappa})\\
					\mu_3(f_1,f_2\epsilon,f_3\epsilon) & = 0,\;\mu_3(f_1,f_2,f_3\epsilon)  = 0,\;\mu_3(f_1,f_2,f_3)  = 0
				\end{aligned}
			\end{equation*}
			One can check that this $\mu_3$ satisfies the required condition for an $A_\infty$ algebra. The resulting Maurer-Cartan equation becomes the Vasiliev equation \eqref{hs_Vasiliev} with $V(X) = c_1X + c_3X\star X\star X,\bar{V}(X) = \bar{c}_1X + \bar{c}_3X\star X\star X$. And this construction can be generalized to arbitrary star product function $V(X)$. 
		\end{remark}

		\subsection{HPT analysis of Vasiliev equation}\label{section-HPT-Vasiliev}
		
		Having known the algebraic structure of Vasiliev equation, we can use techniques introduced in Section \ref{Sec_HPT} to analyze it. The process of reduction of Vasiliev equation to physical degree of freedom is nothing but applying homological perturbation theory. The DGLA or $L_\infty$ structure of Vasiliev equation is transferred to an $L_\infty$ structure on $\mathfrak{hs}[\epsilon]$, which gives us a solution to the deformation problem, or equivalently a full nonlinear unfolded HS equation. Similar reduction process is analyzed in \cite{Sezgin:2000hr,Sezgin:2002ru} and is referred to as curvature expansion therein. Our method using homological perturbation theory gives a compact and clean formula calculating all order vertex of the unfolded higher spin equation, and can reveal some hidden structure in Vasiliev equation.
		
		Firstly, the $Z$ dependent part of the algebra $\mathcal{A}^\bullet$ is the usual de Rham complex. Poincare lemma implies that we can naturally identify 
		$$
		\mathfrak{hs}[\epsilon] = H^{\bullet}(\mathcal{A}^\bullet,d_Z).
		$$
		We have the following homotopy equivalence data
		\begin{equation}\label{hs_Vasiliev}
			h\curved(\mathcal{A}^{\bullet},d_Z)\overset{p}{\underset{i}\rightleftarrows} (\mathcal{H},0)
		\end{equation}
		where we denote $\mathcal{H} = \mathfrak{hs}[\epsilon]$. $i$ is the natural inclusion, $p$ is the projection:
		\begin{equation*}
			p(f(Y,Z)) = f(Y,0),\;\text{ and  }p(\alpha_{\mathcal{I}}dZ^{\mathcal{I}}) = 0,\;\text{ for }|\mathcal{I}| \geq 1
		\end{equation*}
		$h$ is the homotopy operator given by Poincare lemma as follows. For $f$ a $0$-form (a function), we have $h(f(Y,Z)) = 0$. For a $q$-form $\alpha = \sum \alpha_{i_1\dots i_1}(Y,Z)dZ^{i_1}\wedge\dots dZ^{i_q}$, we have
		\begin{equation}
			h(\alpha) = -q \sum\limits_{i_1, \cdots, i_q} Z^{i_1}\int_0^1dt t^{q-1}\alpha_{i_1\dots i_q}(Y,tZ)dZ^{i_2}\wedge\dots dZ^{i_q}
		\end{equation}
		It satisfies
		\begin{equation*}
			dh+hd = i\circ p - 1
		\end{equation*}
		and
		\begin{equation*}
			h\circ i = 0,\; p\circ h = 0, \;   h\circ h  = 0.
		\end{equation*}
		Therefore \eqref{hs_Vasiliev} gives us a SDR data. Note that the choice of this homotopy equivalence data is not unique. Here we use the conventional choice to illustrate the HPT techniques. See Discussion at the end for further remarks at this point.

		Construction of Chapter 3 can be applied, hence we naturally have the following SDR data on the tensor coalgebras
		\begin{equation*}
			T^ch\curved (T^c({\mathcal{A}^{\bullet}}[1]),D_Z)\overset{T^cp}{\underset{T^ci}\rightleftarrows} (T^c(\mathcal{H}[1]),0)
		\end{equation*}
		where
		$$
		T^ci = \sum_{n}i^{\otimes n}, \quad T^cp = \sum_{n}p^{\otimes n}, \quad D_Z = \sum_n\sum_{j = 0}^{n-1}\mathds{1}^{\otimes j}\otimes d_Z \otimes \mathds{1}^{\otimes(n-1-j)}.
		$$
		$T^ch$ is defined by
		\begin{equation*}
			T^ch = \sum_{n}\sum_j\mathds{1}^{\otimes j}\otimes h \otimes (i\circ p)^{\otimes(n-1-j)}.
		\end{equation*}
		The suspension is implicitly assumed in our formula. We denote $\mu$ the multiplication (star product) in $\mathcal{A}^{\bullet}$. It defines a coderivative $D_2$ on $T^c(\mathcal{A}^{\bullet}[1])$ and satisfy $(D_Z+D_2)^2 = 0$. By homological perturbation lemma we have the following data
		\begin{equation*}
			H'\curved (T^c({\mathcal{A}^{\bullet}}[1]),D_Z+D_2)\overset{P'}{\underset{I'}\rightleftarrows} (T^c(\mathcal{H}[1]),Q)
		\end{equation*}
		where
		\begin{equation}
			\begin{aligned}
				Q &= T^cp(1 - D_2T^ch)^{-1}D_2T^ci.\\
			\end{aligned}
		\end{equation}
		By the definition of $h$, $Q$ vanishes on the subalgebra $\mathfrak{hs}[\epsilon]$. And for $a,b \in \mathfrak{hs}[\epsilon]$, $\mu(a,b) = a\star b \in \mathfrak{hs}[\epsilon]$. Therefore $T^chD_2T^ci = 0$, and we find that $Q = T^cpD_2T^ci$. This shows that $Q$ is defined exactly by the usual star product on $\mathfrak{hs}[\epsilon]$.
		
		We conclude that the transferred structure is just $(\mathfrak{hs}[\epsilon],\star)$, which is equivalent to the Vasiliev equation with $V = \bar{V} = 0$ reducing to the undeformed HS equation, as expected.
		
		The next step is to consider adding the linear term of $V$ in Vasiliev equation, or equivalently adding the differential $d_\epsilon$. Consider $d_\epsilon$ as a perturbation of $d_Z$ in the homotopy equivalence data \eqref{hs_Vasiliev}, we have the following perturbed SDR data
		\begin{equation}
			h_\epsilon\curved(\mathcal{A}^{\bullet},d_Z+d_\epsilon)\overset{p_\epsilon}{\underset{i_\epsilon }\rightleftarrows} (\mathcal{H},0)
		\end{equation}
		where we used the fact that the new differential $\partial = p(1 - d_\epsilon h)^{-1}d_\epsilon i$ on $\mathcal{H}$ vanish since $pd_\epsilon = 0$. We have the following
		\begin{equation*}
			\begin{aligned}
				h_\epsilon &= h+ h(1 - d_\epsilon h)^{-1}d_\epsilon h = h+ hd_\epsilon h\\
				p_\epsilon &= p + p(1 - d_\epsilon h)^{-1}d_\epsilon h = p\\
				i_\epsilon& = i + h(1 - d_\epsilon h)^{-1} d_\epsilon i = i+ hd_\epsilon i
			\end{aligned}
		\end{equation*}
		We then apply the tensor construction and find the following data
		\begin{equation*}
			T^ch_\epsilon \curved (T^c({\mathcal{A}^{\bullet}}[1]),D_Z+ D_\epsilon)\overset{T^cp}{\underset{T^ci_\epsilon}\rightleftarrows} (T^c(\mathcal{H}[1]),0)
		\end{equation*}
		where $D_\epsilon = \sum_n\sum_{j = 0}^{n-1}\mathds{1}^{\otimes j}\otimes d_{\epsilon} \otimes \mathds{1}^{\otimes(n-1-j)}$. We omit the suspension map to save the notation as before (the suspension is essential when we calculate the vertex, which gives us the right sign). We perturb the above data by the differential $D_2$, and find that
		\begin{equation*}
			H''\curved (T^c({\mathcal{A}^{\bullet}}[1]),D_1+D_2+ D_\epsilon)\overset{P''}{\underset{I''}\rightleftarrows} (T^c(\mathcal{H}[1]),\mathcal{D})
		\end{equation*}
		where the new differential $\mathcal{D}$ on $T^c(\mathcal{H}[1])$ is given by
		\begin{equation*}
			\mathcal{D} = T^cp(1 - D_2 T^ch_\epsilon)^{-1} D_2 T^c i_\epsilon.
		\end{equation*}
		This differential $\mathcal{D}$ gives us a set of map $\mathcal{D}_k : \mathcal{H}[1]^{\otimes k} \to \mathcal{H}[1],\; k\geq 2$ as follows
		\begin{equation}
			\mathcal{D}_k = \text{proj}_{\mathcal{H}[1]}T^cp(D_2 T^ch_\epsilon)^{k-2}  D_2 T^c i_\epsilon.
		\end{equation}
		They define a set of map $m_k = (-1)^{k(k-1)/2}s^{-1}\circ \mathcal{D}_k \circ s^{\otimes k}$ that satisfy the $A_\infty$ relation.
		
		Calculation shows that $m_2 = p\mu(i\otimes i)$ is the star product on $\mathfrak{hs}[\epsilon]$, therefore the vertices $\mathcal{V}(w,w) = w\star w,\;\mathcal{V}(w,C) = w\star C - C\star\pi(w) $ as we want. The higher terms $m_k,\; k\geq 3$ then encode all higher vertices $\mathcal{V}(\dots)$. For example, we easily find that $m_3$ is
		\begin{equation*}
			\begin{aligned}
				m_3 & = p\mu (h\mu \otimes \mathds{1})(\mathds{1}\otimes h
				d_\epsilon\otimes \mathds{1} ) \\
				& + p \mu (h \mu\otimes\mathds{1})(hd_{\epsilon}\otimes \mathds{1}\otimes \mathds{1} ) \\
				& - p\mu (\mathds{1}\otimes h\mu) (\mathds{1}\otimes\mathds{1}\otimes h
				d_\epsilon)\\
				& - p\mu (\mathds{1}\otimes h\mu)(\mathds{1}\otimes h
				d_\epsilon\otimes \mathds{1} )
			\end{aligned}
		\end{equation*}
		We can denote the above four vertex by $\bullet(\bullet(\bullet)),\;\bullet((\bullet)\bullet),\; (\bullet(\bullet))\bullet,\;((\bullet)\bullet)\bullet$. They can be depicted by trees as
		\begin{center}
			\begin{tikzpicture}[grow' = up]
			\tikzstyle{level 1}=[sibling distance=8mm,level distance = 2mm]
			\tikzstyle{level 2}=[sibling distance=8mm,level distance = 7mm]
			\tikzstyle{level 3}=[sibling distance=8mm,level distance = 7mm]
			\tikzstyle{level 4}=[sibling distance=8mm,level distance = 7mm]
			\tikzstyle{level 5}=[sibling distance=8mm,level distance = 2mm]
			\coordinate
			node {$p$}child {edge from parent[draw=none]
				child { child{node{$i$}} child{	
						child{child{node{$i$} edge from parent[draw=none]} }    child{child{node{$hd_ki$} edge from parent[draw=none]} }
						edge from parent node[right] {$h$} }
				}
			};
			\node at (-1.7,0.5) {$\bullet(\bullet(\bullet)) = -$};
			\end{tikzpicture}
			\begin{tikzpicture}[grow' = up]
			\tikzstyle{level 1}=[sibling distance=8mm,level distance = 2mm]
			\tikzstyle{level 2}=[sibling distance=8mm,level distance = 7mm]
			\tikzstyle{level 3}=[sibling distance=8mm,level distance = 7mm]
			\tikzstyle{level 4}=[sibling distance=8mm,level distance = 7mm]
			\tikzstyle{level 5}=[sibling distance=8mm,level distance = 2mm]
			\coordinate
			node {$p$}child {edge from parent[draw=none]
				child { child{node{$i$}} child{	
						child{child{node{$hd_ki$} edge from parent[draw=none]}}    child{child{node{$i$} edge from parent[draw=none]}}
						edge from parent node[right] {$h$} }
				}
			};
			\node at (-1.7,0.5) {$\bullet((\bullet)\bullet) = -$};
			\end{tikzpicture}
			\begin{tikzpicture}[grow' = up]
			\tikzstyle{level 1}=[sibling distance=8mm,level distance = 2mm]
			\tikzstyle{level 2}=[sibling distance=8mm,level distance = 7mm]
			\tikzstyle{level 3}=[sibling distance=8mm,level distance = 7mm]
			\tikzstyle{level 4}=[sibling distance=8mm,level distance = 7mm]
			\tikzstyle{level 5}=[sibling distance=8mm,level distance = 2mm]
			\coordinate
			node {$p$}child {edge from parent[draw=none]
				child { child{	
						child{child{node{$i$} edge from parent[draw=none]} }    child {child{node{$hd_ki$} edge from parent[draw=none]} }
						edge from parent node[left] {$h$} }  child{node{$i$}}
				}
			};
			\node at (-1.5,0.5) {$(\bullet(\bullet))\bullet =$};
			\end{tikzpicture}
			\begin{tikzpicture}[grow' = up]
			\tikzstyle{level 1}=[sibling distance=8mm,level distance = 2mm]
			\tikzstyle{level 2}=[sibling distance=8mm,level distance = 7mm]
			\tikzstyle{level 3}=[sibling distance=8mm,level distance = 7mm]
			\tikzstyle{level 4}=[sibling distance=8mm,level distance = 7mm]
			\tikzstyle{level 5}=[sibling distance=8mm,level distance = 2mm]
			\coordinate
			node {$p$}child {edge from parent[draw=none]
				child { child{	
						child{child{node{$hd_ki$} edge from parent[draw=none]} }    child{child{node{$i$} edge from parent[draw=none]} }
						edge from parent node[left] {$h$} }  child{node{$i$}}
				}
			};
			\node at (-1.5,0.5) {$((\bullet)\bullet)\bullet =$};
			\end{tikzpicture}
		\end{center}
		Explicitly, for $f_1,f_2,f_3\in\mathfrak{hs}$, we have
		\begin{equation*}
			\begin{aligned}
				m_3(f_1,f_2,f_3\epsilon) & = -p[f_1\star h(f_2\star h(f_3\star \mathcal{K}))]\\
				m_3(f_1,f_2\epsilon,f_3) & = p[h(f_1\star h(f_2\star\mathcal{K}))\star f_3 ] - p[f_1\star h (h(f_2\star\mathcal{K})\star f_3) ]\\
				m_3(f_1\epsilon,f_2,f_3) & =  p[h(h(f_1 \star\mathcal{K})\star f_2)\star f_3 ]\\
				m_3(f_1,f_2\epsilon,f_3\epsilon)&=  p[h(f_1\star h(f_2\star\mathcal{K}))\star f_3 -   f_1\star h(h(f_2\star\mathcal{K})\star f_3 )  + f_1\star h(f_2\star \pi(h(f_3\star\mathcal{K} ))  )  ]\epsilon\\
				m_3(f_1\epsilon,f_2,f_3\epsilon) &= p[h(h(f_1\star\mathcal{K}) \star f_2 )\star f_3 + f_1\star \pi(h(f_2\star h(f_3\star\mathcal{K}) )   )]\epsilon\\
				m_3(f_1\epsilon,f_2\epsilon,f_3) & = p[ h(h(f_1\star \mathcal{K})\star f_2) \star \pi(f_3) - h(f_1\star\pi(h(f_2\star\mathcal{K})))\star \pi(f_3 )  + f_1\star \pi(h( h(f_2\star\mathcal{K})\star f_3 )  ) ]\epsilon
			\end{aligned}
		\end{equation*}
		Anti-symmetrization of $m_3$ gives us the interaction vertices $\mathcal{V}(w,w,C),\mathcal{V}(w,C,C)$ of higher spin equation:
		\begin{equation}
			\begin{aligned}
			\mathcal{V}(w,w,C) =& -c_1\left[ w,z^\alpha\int_0^1 dt_2\left[w,z_\alpha\int dt_1 t_1(C\star \varkappa)_{Z\to t_1Z}\right]_{Z\to t_2Z} \right]_{Z=0} \\
				 &- \bar{c}_1\left[ w,\bar{z}^{\dot{\alpha}}\int_0^1 dt_2\left[w,\bar{z}_{\dot{\alpha}}\int dt_1 t_1(C\star \bar{\varkappa})_{Z\to t_1Z}\right]_{Z\to t_2Z} \right]_{Z=0}\\
				\mathcal{V}(w,C,C) =& -c_1\left[ C,z^\alpha\int_0^1 dt_2\left[w,z_\alpha\int dt_1 t_1(C\star \varkappa)_{Z\to t_1Z}\right]_{Z\to t_2Z} \right]^{\pi}_{Z=0} \\
				& +c_1\left[ w,z^\alpha\int_0^1 dt_2\left[C,z_\alpha\int dt_1 t_1(C\star \varkappa)_{Z\to t_1Z}\right]^{\pi}_{Z\to t_2Z} \right]_{Z=0} \\
				&- \bar{c}_1\left[ C,\bar{z}^{\dot{\alpha}}\int_0^1 dt_2\left[w,\bar{z}_{\dot{\alpha}}\int dt_1 t_1(C\star \bar{\varkappa})_{Z\to t_1Z}\right]_{Z\to t_2Z}\right]^{\pi}_{Z=0}\\
			 &+ \bar{c}_1\left[ w,\bar{z}^{\dot{\alpha}}\int_0^1 dt_2\left[C,\bar{z}_{\dot{\alpha}}\int dt_1 t_1(C\star \bar{\varkappa})_{Z\to t_1Z}\right]^{\pi}_{Z\to t_2Z}\right]_{Z=0}\\
			\end{aligned}
		\end{equation}
		where we define the twisted bracket $[-,-]^{\pi}$ as
		\begin{equation}
		[a,b]^{\pi} = a\star\pi(b) - b\star a
		\end{equation}
		Note that the vertex satisfies $\mathcal{V}(w,C,C) \sim C\epsilon \frac{\delta}{\delta w}\mathcal{V}(w,w,C)$. This is also observed in \cite{Vasiliev:1988sa} and generalize to higher vertices.
		\subsection{Tree description of all vertices}
		What we have done in the last section is to transfer the DGA structure of Vasiliev equation to its cohomology. As we discussed in Section \ref{Sec_tree}, the formula obtained from HPT has a "Feynman diagram" expansion.
		\begin{equation*}
			m_n = \sum_{T\in PBT_{n}}(-1)^{\vartheta(T)} m_T,\; \forall n \geq 2
		\end{equation*}
		Recall that the $PBT_n$ stands for the set of planar binary rooted trees with $n$ leaves. The map $m_T$ is obtained by putting $i_\epsilon$ on the leaves, $\mu$ on the vertices, $h_\epsilon$ on the internal edges and $p$ on the root. We draw it as:
		
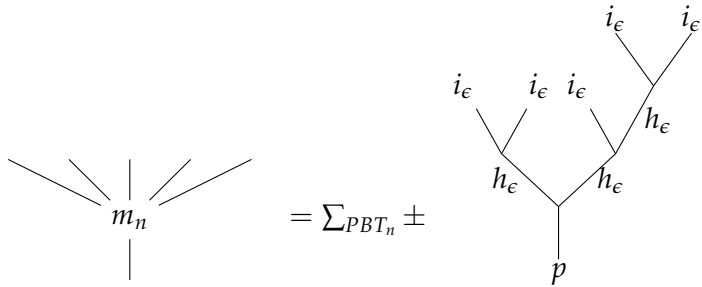
\begin{center}
			\begin{tikzpicture}[grow' = up]
			\tikzstyle{level 1}=[level distance = 2mm]
			\tikzstyle{level 2}=[sibling distance=8mm,level distance = 8mm]
			\tikzstyle{level 3}=[sibling distance=8mm,level distance = 8mm]
			\coordinate
			child {
				edge from parent[draw=none] child {
					node{$m_n$}
					child { child {edge from parent[draw=none]}}  child { child {edge from parent[draw=none]}} child { child {edge from parent[draw=none]}} child { child {edge from parent[draw=none]}} child { child {edge from parent[draw=none]}}
				}
			};
			\node at(3,1) {$ = \sum_{PBT_n}\pm$};
			\end{tikzpicture}	
			\begin{tikzpicture}[grow' = up]
			\tikzstyle{level 1}=[sibling distance=10mm,level distance = 2mm]
			\tikzstyle{level 2}=[sibling distance=10mm,level distance = 7mm]
			\tikzstyle{level 3}=[sibling distance=15mm,level distance = 7mm]
			\tikzstyle{level 4}=[sibling distance=10mm,level distance = 9mm]
			\tikzstyle{level 5}=[sibling distance=10mm,level distance = 7mm]
			\tikzstyle{level 6}=[sibling distance=10mm,level distance = 2mm]
			\coordinate
			node {$p$}child {edge from parent[draw=none]
				child {
					child {	
						child {
							node{$i_\epsilon$}}  child {node{$i_\epsilon$}} edge from parent node[left] {$h_\epsilon$} }    child {	
						child {
							node{$i_\epsilon$}}  child {	
							child {
								child{node{$i_\epsilon$} edge from parent[draw=none]}}  child { child{node{$i_\epsilon$} edge from parent[draw=none]}} edge from parent node[right] {$h_\epsilon$} } edge from parent node[right] {$h_\epsilon$} }
				}
			};
			\end{tikzpicture}
		\end{center}
		For example, for $T$ as in the above figure, $m_T$ is
		\begin{equation*}
			m_T = p\mu(h_\epsilon\mu \otimes h_\epsilon \mu)(\mathds{1}^{\otimes 3}\otimes h_\epsilon \mu)i_\epsilon^{\otimes 5}.
		\end{equation*}
		
		In this section, we mostly follow \cite{Markl} for the description of the sign $\vartheta(T)$ appearing in the formula. First we need to introduce some notation. For a vertex $v$ of $T$, we call the inputs of $v$ as the set of incoming edges connected to $v$. For planar binary tree, every vertex has 2 inputs, and we label them by $1$ and $2$ from left to right. For a vertex $v$, we denote $r_i,\;i = 1,2$ the number of leaves of $T$ such that the unique path from this leave to root contain the $i$-th input of $v$. We define the function
		\begin{equation*}
			\vartheta_T(v) = \vartheta(r_1,r_2) = r_{1}(r_2 + 1)
		\end{equation*}
		Then the sign is
		$$
		\vartheta(T) = \sum_{v \in V_T}\vartheta_T(v).
		$$
		For example, considering the following $T$
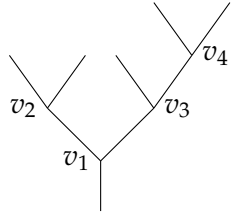
		\begin{center}
			\begin{tikzpicture}[grow' = up]
			\tikzstyle{level 1}=[sibling distance=10mm,level distance = 8mm]
			\tikzstyle{level 2}=[sibling distance=14mm,level distance = 7mm]
			\tikzstyle{level 3}=[sibling distance=10mm,level distance = 7mm]
			\tikzstyle{level 4}=[sibling distance=10mm,level distance = 7mm]
			\tikzstyle{level 5}=[sibling distance=10mm,level distance = 7mm]
			\node (root) {~}
			child {
				child {	
					child   child  }    child {	
					child   child {	
						child   child   } }
			};
			\node at (root-1)[left] {$v_1$};
			\node at (root-1-1)[left] {$v_2$};
			\node at (root-1-2)[right] {$v_3$};
			\node at (root-1-2-2)[right] {$v_4$};
			\end{tikzpicture}
		\end{center}
		There are four vertices $v_1,v_2,v_3,v_4$. We find that for $v_1$, $(r_1,r_2)=(2,3)$ $\vartheta_T(v_1) = 2\cdot 4$, for $v_2$, $(r_1,r_2)=(1,1)$ $\vartheta_T(v_2) = 1\cdot 2$, for $v_3$, $(r_1,r_2)=(1,2)$ $\vartheta_T(v_3) = 1\cdot 3$, and for $v_4$, $(r_1,r_2)=(1,1)$ $\vartheta_T(v_2) = 1\cdot 2$. Therefore $\vartheta(T) = 1\mod 2$.

		Since $i_\epsilon = i + hd_\epsilon i$, and $h_{\epsilon} = h+ hd_\epsilon h$, we can further expand every $m_T$ by replacing each $i_\epsilon$ by either $i$ or $hd_\epsilon i$, and $h_{\epsilon}$ by either $h$ or $hd_\epsilon h$ and then sum them together. To make this precise, for each tree $T$, we define the set
		\begin{equation*}
			\mathcal{S}_T = \{f:E_T\backslash \{\text{root}\}\to \{1,2\} \}
		\end{equation*}
		where $E_T$ refers to the set of edges of $T$, and we denote $\text{root}$ the only edge connected with root. For each assignment $f$, we define a map $m_{T,f}$ as follows. For $e$ a leave, we put $i$ on it if $f(e) = 1$, and we put $hd_\epsilon i$ on it if $f(e) = 2$. For $e$ an internal edge, we put $h$ on it if $f(e) = 1$, and we put $hd_\epsilon h$ on it if $f(e) = 2$. All vertices are still assigned with $\mu$, and the root is assigned with $p$. Then every $m_T$ has the following expansion:
		\begin{equation*}
			m_T = \sum_{f\in \mathcal{S}_T}m_{T,f}
		\end{equation*}
		However, note that many $m_{T,f}$ actually equal zero. For example, the following maps all equal to zero.
		
		\begin{figure}[h!]
			\centering
			\begin{subfigure}[h!]{0.3\textwidth}
				\begin{tikzpicture}[grow' = up]
				\tikzstyle{level 1}=[sibling distance=10mm,level distance = 8mm]
				\tikzstyle{level 2}=[sibling distance=16mm,level distance = 7mm]
				\tikzstyle{level 3}=[sibling distance=10mm,level distance = 9mm]
				\tikzstyle{level 4}=[sibling distance=10mm,level distance = 9mm]
				\tikzstyle{level 5}=[sibling distance=10mm,level distance = 9mm]
				\node (root) {$p$}
				child {
					child {	
						child{node{$hd_\epsilon i$}}  child {node{$i$}}  edge from parent node[left] {$h$}}    child {	
						child {node{$hd_\epsilon i$}}  child {	
							child{node{$hd_\epsilon i$}}   child{node{$hd_\epsilon i$}}   edge from parent node[right] {$h$}} edge from parent node[right] {$h$}}
				};
				\node at (root-1)[left] {$v_1$};
				\node at (root-1-1)[left] {$v_2$};
				\node at (root-1-2)[right] {$v_3$};
				\node at (root-1-2-2)[right] {$v_4$};
				\end{tikzpicture}
				\caption{~}
				\label{fig:vanish_tree1}
			\end{subfigure}
			~
			\begin{subfigure}[h]{0.3\textwidth}
				\begin{tikzpicture}[grow' = up]
				\tikzstyle{level 1}=[sibling distance=10mm,level distance = 8mm]
				\tikzstyle{level 2}=[sibling distance=16mm,level distance = 7mm]
				\tikzstyle{level 3}=[sibling distance=10mm,level distance = 9mm]
				\tikzstyle{level 4}=[sibling distance=10mm,level distance = 9mm]
				\tikzstyle{level 5}=[sibling distance=10mm,level distance = 9mm]
				\node (root) {$p$}
				child {
					child {	
						child{node{$i$}}  child {node{$i$}}  edge from parent node[left] {$h$}}    child {	
						child {node{$i$}}  child {	
							child{node{$i$}}   child{node{$i$}}   edge from parent node[right] {$h$}} edge from parent node[right] {$h$}}
				};
				\node at (root-1)[left] {$v_1$};
				\node at (root-1-1)[left] {$v_2$};
				\node at (root-1-2)[right] {$v_3$};
				\node at (root-1-2-2)[right] {$v_4$};
				\end{tikzpicture}
				\caption{~}
				\label{fig:vanish_tree2}
			\end{subfigure}
			\caption{~}
		\end{figure}
		The first one equals to zero because we will get a $1$-form before the projection map $p$. The second one equals to zero because $h$ is zero when restricted on $\mathfrak{hs}$.
		
		We can find a rule to determine if an assignment $f$ gives us a zero map $m_{T,f}$. First define the form degree of map by $\deg' i = \deg' \mu = \deg' p = 0$, $\deg' d_\epsilon = 2$ and $\deg' h = -1$. We also define a partial ordering on the set $V_T\cup E_T$: $x \preccurlyeq y$ if and only if $y$ is contained in the unique path from $x$ to the root. Given an assignment $f$, we denote $\phi_f(x)$ the map assigned on the vertex or edge $x$. Since each vertex is assigned with the multiplication we have $\phi_f(v) = \mu$ and $\deg'\phi_f(v) = 0$. Then we define the form degree of a vertex $v$ by
		\begin{equation*}
			\deg_f v = \sum_{x \preccurlyeq v,x \in E_T}\deg' \phi_f(x)
		\end{equation*}
		For example, for the map of fig \ref{fig:vanish_tree1}, we have four vertex $v_1,v_2,v_3,v_4$, and we can calculate that $\deg_f v_4 = 2,\; \deg_f v_3 = 2,\; \deg_f v_2 = 1,\; \deg_f v_1 =1$. This map equals to zero just because $\deg_f v_1 =1$, since after projection $p$ it gives us $0$. For the map of fig \ref{fig:vanish_tree2}, we have $\deg_f v_4 = 0,\; \deg_f v_3 = -1, \deg_f v_2 = 0, \deg_f v_1 = -3$. This map equals zero because $\deg_f v_4 = 0,\; \deg_f v_2 = 0$, for the homotopy $h$ after them.
		
		We summarize the rule as follows. For a tree $T$, an assignment $f$ is called admissible if $\deg_f v \geq 1$ for all $v$ not connected to the root and $\deg_f v_1 = 0$ where $v_1$ is the unique vertex connected to the root. We denote
		\begin{equation*}
			\Tilde{\mathcal{S}}_T = \{f:E_T\backslash \{\text{root}\}\to \{1,2\},\; f\text{ admissible } \}
		\end{equation*}
		Then we have the following formula for the $A_\infty$ structure
		\begin{equation}\label{tree_formula_2}
			\boxed{m_n = \sum_{T\in PBT_{n}}(-1)^{\vartheta(T)} \sum_{f \in \Tilde{\mathcal{S}}_T} m_{T,f},\; \forall n \geq 2}
		\end{equation}
		
		Expanding this formula for $m_3$, we find exactly the same result as we calculated using homological perturbation theory.
		
		\begin{remark}
			The tree construction easily generalizes to the case when the function $V$ of Vasiliev equation contain cubic or higher order term. As we discussed previously, they correspond to an $A_\infty$ structure with non vanishing higher product $\mu_k,\; k\geq 3$. There are still tree description of the transferred structure. We need to consider summing over all rooted planar trees, and we assign $\mu_k$ to vertex with $k$ inputs edges.
		\end{remark}

		\section{Relation to a topological quantum mechanics model}
		
		The appearance of Weyl algebra in higher spin theory suggests that it might be related to quantum mechanics. Indeed, we will show in this section that there is a topological quantum mechanics model hidden (at least at first order) in the nonlinear higher spin theory reduced from Vasiliev equation. The general idea is that, as we explained previously, the construction of nonlinear higher spin theory is a deformation problem. Generally, the first order deformation of an algebra is controlled by its Hochschild cohomology. We will explain that in the topological quantum mechanics model, the Hochschild cocycle condition, or equivalently the consistence of higher spin equation to first order, is implemented through the quantum master equation.
		
		\subsection{TQM and Hochschild homology}
		\label{TQM&Hoch}
		We will describe a toy model of topological quantum mechanics \cite{Si-index} to explain its relation with the vertex of Vasiliev equation. To illustrate the general idea, we consider first an easier model for Hochschild cocycle of the Weyl algebra.
		
		\paragraph{FFS cocycle}
		It is relatively easy to calculate the dimensions of Hochschild (co)homology of Weyl algebra, and it is known that
		\begin{equation}
			HH^{j}(A_n) =\begin{cases}
				\mathbb{C},& j = 2n\\
				0,&\text{otherwise}
			\end{cases}
		\end{equation}
		However, it is much more difficult to find an explicit expression for the nontrivial $2n$-cocycle $\tau_{2n}$\cite{Feigin}, which is related to the formality theorem for Hochschild chains \cite{Kontsevich:1997vb,Shoikhet:2000gw}.
		
		We first explain the FFS formula for $\tau_{2n}$\cite{Feigin}.  Let us fix the standard symplectic structure
		\begin{equation*}
			\omega = \sum_{i = 1}^{n}dy^{2i -1}\wedge dy^{2i}.
		\end{equation*}
		Let $\Delta_{2n}$ be the standard $2n$-simplex:
		\begin{equation*}
			\Delta_{2n} = \{0 = u_0\leq u_1 \leq \dots \leq u_{2n}\leq 1\}.
		\end{equation*}
		We denote partial derivatives of the $k$-th component of the $i$-th variable $y_i$ by
		\begin{equation*}
			(p_i)_{k} = \frac{\partial}{\partial (y_i)^k}, \quad i = 0,1,\dots,2n
		\end{equation*}
		and write
		\begin{equation*}
			\alpha_{ij} = \omega^{kl}(p_i)_k(p_j)_l.
		\end{equation*}
		Regarding $(p_i)_{k},1 \leq i,k \leq 2n$ as a matrix, we denote $\pi_{2n}$ to be the determinate $\pi_{2n} = \det((p_i)_k)$. Then the cocycle $\tau_{2n}$ has the following form:
		\begin{equation}\label{Hosch_Weyl}
			\tau_{2n}(f_0,\dots,f_{2n}) = \pi_{2n}\int_{\Delta_{2n}}d^{2n}u \exp\left[\sum_{0\leq i <j \leq 2n}\hbar(u_i-u_j - \frac{1}{2})\alpha_{ij}\right]f_0(y_0)\cdots f_{2n}(y_{2n})|_{y_0=\cdots y_{2n} = 0}
		\end{equation}
		
		In \cite{Feigin}, this formula is found by simplifying integrals over configuration spaces that appear in the formality conjecture. We mention that similar formula was found in \cite{Vasiliev:1988sa} in the special case of $A_2$. In the following part of this section, we will provide a topological quantum mechanics interpretation for this cocycle.
		
		\paragraph{Field content and Lagrangian}  The Hochschild cocyle $\tau_{2n}$ can be obtained by an explicit path integral for specified topological observables in topological quantum mechanics model. This follows from the construction in \cite{Si-index} that we briefly describe here. 
		
		We consider the standard symplectic space $(\mathbb{R}^{2n},\omega)$ where
		\begin{equation*}
			\omega = \sum_{i = 1}^n dx^{i}\wedge dp_i.
		\end{equation*}
		
		We associate a topological quantum mechanics model as follows: the fields are
		\begin{equation*}
			\mathcal{E} = \Omega^{\bullet}(I)\otimes \mathbb{R}^{2n}
		\end{equation*}
		Here $I$ is a one-dimensional manifold parametrizing the time $t$. The fields are given by $2n$ copies of differential forms on $I$. We can describe the fields in terms of components by
		\begin{equation*}
			\mathbb{X}^i(t) = x^i(t) + \xi^i(t)dt,\quad \mathbb{P}_i(t) = p_i(t) + \eta_i(t)dt
		\end{equation*}
		where $x^i(t),p_i(t)$ are bosons, $\xi^i(t),\eta_i(t)$ are fermions and anti-fields of the bosons. We assign the cohomology degree (ghost number) by
		\begin{equation*}
			\deg(x^i) = \deg(p^i) = 0,\quad \deg(\xi_i) = \deg(\eta_i) = -1.
		\end{equation*}
		We introduce the BRST operator $Q$ of degree $1$ representing the de Rham differential:
		\begin{equation*}
			Qx^i = 0,\; Q\xi^i = \partial_t x_i,\; Qp_i = 0,\; Q\eta_i = \partial_t p_i.
		\end{equation*}
		The action functional in the BRST-BV formalism is the free one
		\begin{equation}
			S =\int_{I}\mathbb{P}_id\mathbb{X}^i = \int_Ip_idx^i
		\end{equation}
		which is manifest BRST invariant. We will consider the case when $I = S^1$.
		
		\paragraph{Gauge fixing condition}
		To perform the path integral in the BRST-BV formalism, we need to choose a super lagrangian subspace
		\begin{equation}
			\mathcal{L} \subset \mathcal{E}.
		\end{equation}
		This lagrangian subspace $\mathcal{L}$ is also called gauge fixing condition. This allows us to compute correlation function for an observable $\mathcal{O}$ by
		\begin{equation*}
			\left\langle \mathcal{O}\right\rangle  = \int_{\mathcal{L}}\mathcal{O}e^{S/\hbar}.
		\end{equation*}
	In particular, if $\mathcal O$ is closed under BV operator, then this value does not depend on continuous deformations of $\mathcal L$. This features quantum gauge invariance in the standard BRST-BV formalism. 	
		
		To specify the gauge fixing condition, we choose the standard flat metric on $S^1$. Let $d^*$ be the adjoint of $d$. Hodge theory gives the decomposition 
		\begin{equation*}
			\Omega^{\bullet}(S^1) = \text{Im}d \oplus \text{Im}d^* \oplus \mathcal{H}(S^1)
		\end{equation*}
	where $\mathcal{H}(S^1)$ is the space of Harmonic forms representing the zero modes. Denote 
	$$
	\mathcal{H}(S^1)^{\perp} = \text{Im}d \oplus \text{Im}d^*.
	$$ 
	Equivalently 
		\begin{equation*}
			\mathcal{H}(S^1) = \{f+gdt\mid f,g \in\mathbb{R}\} = \mathbb{R}[dt],\quad \;\mathcal{H}(S^1)^{\perp} = \{f(t)+g(t)dt\mid \int_{S^1}f = \int_{S^1}g = 0\}.
		\end{equation*}
		We choose the gauge fixing condition $\mathcal{L}$ in such a way that:
		\begin{equation*}
			\begin{aligned}
				\mathcal{L}_1 & :=	\mathcal{L}\cap (\mathcal{H}(S^1)^{\perp}\otimes \mathbb{R}^{2n})= \{\xi^i(t) = \eta_i(t) = 0\}=\Im d^*. \\
				\mathcal{L}_2 & :=\mathcal{L}\cap(\mathcal{H}(S^1)\otimes \mathbb{R}^{2n}) = \{x^i = p_i = 0\}\\
				\mathcal L&=\mathcal L_1\oplus \mathcal L_2
			\end{aligned}
		\end{equation*}
		In other words
		\begin{equation*}
			\mathcal{L} = \{x^i(t) + \xi^idt,p_i(t)+\eta_idt\mid \int_{S^1}x^i(t) = \int_{S^1}p_i(t) = 0\}.
		\end{equation*}
		The choice of the lagrangian subspace $\mathcal{L} = \mathcal{L}_1\oplus \mathcal{L}_2$ tells us how to do path integral. Integration over $\mathcal{L}_1$ can be computed by Feynman diagrams with the propagator that we will soon calculate; integration over  $\mathcal{L}_2$ is implemented by a Berezin integral $\int d^n\xi d^n\eta$.
	
		The nontrivial propagator on the subspace $\mathcal{L}_1$ is given by
		\begin{equation*}
			\left\langle x^i(t_1)p_j(t_2) \right\rangle = \delta^i_j P(t_1,t_2)
		\end{equation*}
		where
		\begin{equation}\label{propagator-formula}
			P(t_1,t_2) = t_1 - t_2 - \frac{1}{2},\;\text{ for } 0 < t_1 - t_2 < 1, t_i \in S^1=\R/\Z.
		\end{equation}
		The propagator is computed as follows (see \cite[Appendix B]{Si-index}). The choice of gauge fixing condition $\mathcal L_1$ implies that the propagator is the kernel of the operator 
		$$
		   d^*{1\over \Box}=\int_0^\infty d^* e^{-u\Box}du
		$$
	where $\Box=dd^*+d^*d$ is the Laplacian on forms, $1\over \Box$ is the Green's operator, and $e^{-u\Box}$ is the heat operator. 
		Recall the heat kernel function on the unit circle $S^1=\R/\Z$ is given by
		\begin{equation*}
			\begin{aligned}
				K_t(\theta_1,\theta_2) &=  \frac{1}{\sqrt{4\pi t}} \sum_{n \in \mathbb{Z}}e ^{-\frac{(\theta_1 - \theta_2 + n)^2}{4t}}= \sum_{n\in\mathbb{Z}} e^{-4\pi^2 n^2 t}e^{2\pi i n (\theta_1 - \theta_2)}.
			\end{aligned}
		\end{equation*}
		The propagator can then be calculated through
		\begin{equation*}
			\begin{aligned}
				P(\theta_1,\theta_2)& = \int_{0}^\infty \frac{\partial}{\partial \theta_1} K_t(\theta_1,\theta_2) dt = \sum_{n\in \mathbb{Z}\backslash\{0\}} \frac{i}{2\pi n} e^{2\pi in(\theta_1 - \theta_2)}.
			\end{aligned}
		\end{equation*}
		This is precisely the Fourier expansion of the function in \eqref{propagator-formula}. 
		
	\paragraph{BV structure}	
		
		The BV anti-bracket is
		\begin{equation*}
			\{x^i(t_1),\eta_j(t_2)\} = \{\xi^i(t_1),p_j(t_2)\} = \delta^i_j\delta(t_1 - t_2).
		\end{equation*}
		The BV operator $\Delta$ is given by the standard form
		\begin{equation*}
			\Delta = \int dt_1dt_2 \delta(t_1 - t_2)\sum_i\left( \frac{\delta}{\delta x^i(t_1)}\frac{\delta}{\delta \eta_i(t_2)} +  \frac{\delta}{\delta \xi^i(t_1)}\frac{\delta}{\delta p_i(t_2)}\right).
		\end{equation*}
		Strictly speaking this is ill-defined due to the UV divergence and requires renormalization. We refer to \cite{Si-index} for a careful treatment of renormalized BV operator, as well as the rigorous meaning for various naive formulae in the discussions below. 
		\paragraph{Topological observables} Given a function $f(x,p)$, we associate the following local observables $f(\mathbb{X}^i(t),\mathbb{P}^i(t))=f^{(0)}(t)+f^{(1)}(t)$ where
		\begin{equation*}
			f^{(0)}(t) = f(x(t),p(t)), \quad f^{(1)}(t) = \sum_{i}(\xi^{i}(t)\partial_{x^i}f(x(t),p(t)) + \eta_i(t)\partial_{p_i}f(x(t),p(t))dt.
		\end{equation*}
		One can easily check that
		\begin{equation*}
			Qf^{(0)} = 0,\;df^{(0)} = Qf^{(1)},\Delta f^{(0)} = \Delta f^{(1)} = 0.
		\end{equation*}
		It follows that
		\begin{equation*}
			(Q + \hbar\Delta)f^{(0)} = 0,\quad (Q + \hbar\Delta) \int_{S^{1}}f^{(1)} = 0.
		\end{equation*}
		Also note that
		\begin{equation*}
			f^{(0)}(t_2) - f^{(0)}(t_1) = (Q + \hbar\Delta)\int_{t_1}^{t_2}f^{(1)}
		\end{equation*}
		which says that the BV homology class of $f^{(0)}(t)$ does not depend on the position of $t$. This is a general property of topological field theory.
		\paragraph{Operator product expansion}
		Using the propagator we can compute the OPE (still denoted by $\star$) for two local observables
		\begin{equation*}
			f^{(0)}(t_1)\star g^{(0)}(t_2) = f^{(0)}(t_1)e^{\hbar P(t_1,t_2)\sum_i(\overleftarrow{\partial}_{x^i}\overrightarrow{\partial}_{p_i} - \overleftarrow{\partial}_{p_i}\overrightarrow{\partial}_{x^i})}g^{(0)}(t_2)
		\end{equation*}
		In particular, we find that
		\begin{equation}
			\lim_{t_1 \to t_2} f^{(0)}(t_1)\star g^{(0)}(t_2)= (f\star g)^{(0)}(t_2)
		\end{equation}
		where $f\star g$ is the standard Moyal star product.
		
		We consider the effect of BV operator on product of observables. First, for the simplest case, we have
		\begin{equation*}
			(Q + \hbar \Delta) (f^{(0)}(t)\int_{S^1}g^{(1)} )= [f,g]_*^{(0)}(t).
		\end{equation*}
		Here we have used OPE to expand the fields (given by the Moyal star product as above) when $(Q+\hbar\Delta)$ brings $f$ and $g$ to the same point. Generally, we have 
		\begin{equation*}
			\begin{aligned}
				(Q &+ \hbar \Delta)\left(\int_{\Delta_{m}}f^{(0)}_0(t_0)f_1^{(1)}(t_1) f^{(1)}_2(t_2) \cdots f_m^{(1)}(t_m) \right)\\
				& = \int_{\Delta_{m-1}}(f_0\star f_1)^{(0)}(t_0) f^{(1)}_2(t_2) \cdots f_m^{(1)}(t_m) ) \\
				& + \sum_{i=1}^{m-1}(-1)^{i} f^{(0)}_0(t_0)f_1^{(1)}(t_1) \cdots (f_i\star f_{i+1})^{(1)}(t_i) \cdots f_m^{(1)}(t_m)\\
				& + (-1)^m\int_{\Delta_{m-1}}(f_m\star f_0)^{(0)}(t_0) f^{(1)}_2(t_2) \cdots f_{m-1}^{(1)}(t_{m-1})
			\end{aligned}
		\end{equation*}
		\paragraph{TQM interpretation of FFS cocycle}
		We define the following map
		\begin{equation}
			\Phi(f_0,\dots,f_{2n}) = \left<\int_{\Delta_{2n}}f^{(0)}_0(u_0)f_1^{(1)}(u_1)f_2^{(1)}(u_2)\cdots f_{2n}^{(1)}(u_{2n})\right>.
		\end{equation}
		This correlation function is computed with our gauge fixing $\mathcal L$ and can be represented by
		
		
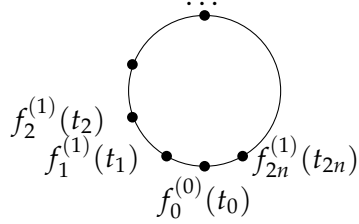
\begin{center}
			\begin{tikzpicture}
			\draw(0,0) circle (1);
			\fill (0,-1) circle (2pt) node[below] {$f^{(0)}_0(t_0)$};
			\fill (-0.5,-0.866) circle (2pt) node[left] {$f_1^{(1)}(t_1)\;\;$};
			\fill (-0.95,-0.35) circle (2pt) node[left] {$f_2^{(1)}(t_2)\;\;$};
			\fill (-0.95,0.35) circle (2pt);
			\fill (0,1) circle (2pt) node[above] {$\dots $};
			\fill (0.5,-0.866) circle (2pt) node[right]{$f_{2n}^{(1)}(t_{2n})$};
			\end{tikzpicture}
		\end{center}

		Let $\partial$ denote the Hochschild differential
		\begin{equation*}
			\partial \Phi(f_0,\dots,f_{2n+1}) = \sum_{i = 0}^{2n}(-1)^{i}\Phi(f_0,\dots,f_i\star f_{i+1},\dots,f_{2n+1}) - \Phi(f_{2n+1}\star f_0,f_1,\dots,f_{2n})
		\end{equation*}
		We see that $\Phi$ is a Hochschild cocycle since
		\begin{equation*}
			\partial \Phi(f_0,\dots,f_{2n+1})  = \left<(Q + \hbar \Delta)\int_{\Delta_{2n+1}}f^{(0)}_0(u_0)f_1^{(1)}(u_1)f_2^{(1)}(u_2)\cdots f_{2n+1}^{(1)}(u_{2n+1})\right> = 0
		\end{equation*}
		Using the propogator we easily calculate that $\Phi$ is precisely the Hoschschild cocycle $\tau_{2n}$ in \eqref{Hosch_Weyl}.
		\begin{equation}
			\tau_{2n}(f_0,\dots,f_{2n}) = \Phi(f_0,\dots,f_{2n})
		\end{equation}
		
		The above constrution gives us the following quantum mechanical interpretation of the Hochschild cohomology of Weyl algebra
		\begin{framed}
			\begin{center}
				Hoschschild cocycle condition $\Leftrightarrow$ BV quantum master equation
			\end{center}
		\end{framed}
		The construction can be generalized, and we will consider a topological quantum mechanics model that calculates vertices in higher spin equation.
		
		\subsection{Vertex $m_3$ in Vasiliev equation}
		In this section we perform the calculation of the vertex $m_3$. We write 
		$$
		u\cdot v = u_\alpha v^\alpha = \epsilon_{\alpha\beta}u^\alpha v^\beta, \quad \bar{u}\cdot \bar{v} = \bar{u}_{\dot{\alpha}} \bar{v}^{\dot{\alpha}} = \epsilon_{\dot{\alpha}\dot{\beta}}\bar{u}^{\dot{\alpha}} \bar{v}^{\dot{\beta}}, \quad U\cdot V = U_A V^A = \epsilon_{AB}U^A V^B.
		$$ 
		For any function $f_1(Y)$, we can write
		\begin{equation*}
			f_1(Y) = e^{-Y\cdot P_1}f_1(Y_1)|_{Y_1 = 0}
		\end{equation*}
		where $P_1 = \partial/\partial Y_1$. In general, we consdier linear operator acting on $n$ functions that can be represented by its symbol as a function of $P_i,1\leq i\leq n $.
		\begin{equation*}
			\Psi(f_1,\dots,f_n)(Y) = \nu(Y,P_1,\dots,P_n)f_1(Y_1)\cdots f_n(Y_n)|_{Y_i = 0}
		\end{equation*}
		For example, the Moyal star product, as a bilinear operator, can be written as
		\begin{equation*}
			(f_1\star f_2)(Y) = e^{\hbar P_1\cdot P_2 - Y\cdot P_1 - Y\cdot P_2}f_1(Y_1)f_2(Y_2)|_{Y_i = 0}
		\end{equation*}
		We also introduce a trace on Weyl algebra
		\begin{equation*}
			\text{tr}(f(Y)) = f(0)
		\end{equation*}
		and define a bilinear pairing $B(\cdot,\cdot)$
		\begin{equation*}
			B(f,g) = \text{tr}(f\star g).
		\end{equation*}
		This pairing allows us to identify a linear operator $\Psi: V^{*\otimes n} \to V^{*}$ with a linear map $V^{*\otimes n+1} \to k$ through
		\begin{equation*}
			\begin{aligned}
				B(f_0,\Psi(f_1,\dots,f_n)) &= \text{tr}(e^{-Y\cdot P_0}\star \nu(Y,P_1,\dots,P_n))f_0(Y_0)f_1(Y_1)f_2(Y_2)|_{Y_i = 0}\\
				& =\nu(-\hbar P_0,P_1,\dots,P_n)f_0(Y_0)f_1(Y_1)f_2(Y_2)|_{Y_i = 0}
			\end{aligned}
		\end{equation*}
		We will find the operator $\nu$ correspond to the vertices of $m_3$.
		
		\paragraph{1. $\bullet(\bullet(\bullet))$}
		Product with Klein operator can be written as
		\begin{equation*}
			f_3(y,\bar{y})\star \varkappa = f_3(-z,\bar{y})e^{-zy/\hbar} = e^{z\cdot p_3 - \frac{1}{\hbar}z\cdot y - \bar{y}\cdot \bar{p}_3} f_3,\quad f_3(y,\bar{y})\star \bar{\varkappa} = f_3(y,-\bar{z})e^{-\bar{z}\cdot\bar{y}/\hbar} = e^{\bar{z}\cdot\bar{p}_3 - \frac{1}{\hbar}\bar{z}\cdot\bar{y} - y\cdot p_3} f_3
		\end{equation*}
		Therefore we have
		\begin{equation*}
			\bullet(\bullet(\bullet))= -p[e^{-yp_1 - \bar{y}\bar{p}_1}\star h(e^{-yp_2-\bar{y}\bar{p}_2}\star h(c_1d^2ze^{z(-\frac{1}{\hbar}y+p_3)-\bar{y}\bar{p}_3} + \bar{c}_1d^2\bar{z}e^{\bar{z}(-\frac{1}{\hbar}\bar{y}+\bar{p}_3) -yp_3 }))] f_1f_2f_3
		\end{equation*}
		Since the star product does not mix $y,\bar{y}$ we can first calculate the following:
		\begin{equation*}
			\begin{aligned}
				&p[e^{-yp_1}\star h(e^{-yp_2}\star h(d^2ze^{z(-\frac{1}{\hbar}y+p_3)}))]\\	
				= &p[e^{-yp_1}\star h(e^{-yp_2}\star(\epsilon_{\alpha\beta} z^\alpha dz^\beta t_0e^{t_0z(-\frac{1}{\hbar}y+p_3)}))]\\
				=&\int_0^1 dt_0 p[e^{-yp_1}\star h(e^{-yp_2}(\epsilon_{\alpha\beta} (z^\alpha + \hbar p_2^\alpha) dz^\beta t_0e^{t_0(z + \hbar p_2)(-\frac{1}{\hbar}y + p_2 +p_3)}))]\\
				= &\int_0^1 dt_0dt_1 p[e^{-yp_1}\star (e^{-yp_2}(\epsilon_{\alpha\beta} (t_1z^\alpha + \hbar p_2^\alpha)z^\beta t_0e^{t_0(t_1z + \hbar p_2)(-\frac{1}{\hbar}y +p_2 +p_3)}))]\\
				=& \int_0^1 dt_1dt_0p[e^{-yp_1}\star (e^{-yp_2}(\epsilon_{\alpha\beta} (  \hbar p_2^\alpha)z^\beta t_0e^{t_0(t_1z + \hbar p_2)(-\frac{1}{\hbar}y + p_2 +p_3)}))]\\
				= &-\hbar^2\int_0^1 dt_1dt_0 p_1\cdot p_2 t_0e^{-yp_1 + (-y + \hbar p_1)p_2 + t_0(\hbar t_1p_1 + \hbar p_2)(-\frac{1}{\hbar}y +p_1 + p_2 +p_3) }
			\end{aligned}
		\end{equation*}
		We make the following change of variable
		\begin{equation*}
			t_0t_1 = 2u_1,\;t_0 = 2u_2,\quad 0<t_0,t_1 < 1 \to 0<u_1<u_2 < \frac{1}{2},\quad dt_0dt_1 = \frac{4}{t_0}du_1du_2
		\end{equation*}
		Then we find
		\begin{equation*}
			\begin{aligned}
				&	p[e^{-yp_1}\star h(e^{-yp_2}\star h(d^2ze^{z(iy+p_3)}))]\\
				= &-4\hbar^2\int du_1du_2 p_1\cdot p_2e^{-yp_1(1 - 2u_1) - yp_2(1 - 2u_2) + \hbar p_1p_2( 1- 2u_2 + 2u_1) +  \hbar p_1p_3 2u_1 + \hbar p_2p_32u_2 }
			\end{aligned}
		\end{equation*}
		Using the bilinear pairing we turn $y \to -\hbar p_0$, the symbol of the above operator can be written as
		\begin{equation*}
			\int du_1du_2 p_1\cdot p_2e^{\hbar 2p_0\cdot p_1P(u_1 - u_0) +\hbar 2p_0\cdot p_2P(u_2-u_0) + \hbar 2p_1\cdot p_2P(u_2 - u_1) +  \hbar 2p_1\cdot p_3P(u_3 - u_1) + \hbar2p_2\cdot p_3P(u_3 - u_2) }
		\end{equation*}
		where we denote $P(u) = \frac{1}{2} - u$, and $u_0 = 0,u_3 = \frac{1}{3}$. Restoring the anti-holomorphic part, we find that the full vertex including the anti-holomorphic variable is given by
		\begin{equation*}
			m_{3,1} := \bullet(\bullet(\bullet)) = 4\hbar^2c_1\int_{\Delta^1} du_1du_2 p_1p_2e^{\hbar\sum_{j<l}2\bar{p}_j\bar{p}_l\bar{P}(u_l - u_j)+\hbar\sum_{j<l}2p_jp_l P(u_l - u_j)}+ c.c
		\end{equation*}
		where we define $\bar{P}(u) = \frac{1}{2}$. The integration is taken over $\Delta^{1} = \{0  = u_0< u_1 < u_2 < u_3 = \frac{1}{2}\}$. The "complex conjugate" is implemented by $c_1 \to \bar{c}_1,\;p_i \to \bar{p}_i$.
		
		\paragraph{2. $(\bullet(\bullet))\bullet$}
		The holomorphic part of this vertex is
		\begin{equation*}
			\begin{aligned}
				&p[h(e^{-yp_1}\star h(d^2ze^{z(-\frac{1}{\hbar}y+p_2)}))\star e^{-yp_3}]\\
				= &-4\hbar^2 \int du_1du_2 p_1p_3e^{-yp_1(1 - 2u_2) - yp_3(1 - 2u_1) + \hbar p_1p_3( 1- 2u_2 - 2u_1) + \hbar p_1p_2 (2u_2) + \hbar p_2p_3(-2u_1) }
			\end{aligned}
		\end{equation*}
		where $0<u_1<u_2<\frac{1}{2}$.
		\paragraph{3. $\bullet((\bullet)\bullet)$}
		\begin{equation*}
			\begin{aligned}
				-&p[e^{-yp_1}\star h(h(d^2ze^{z(-\frac{1}{\hbar} y+p_2)})\star e^{-yp_3})]\\
				= &-4\hbar^2 \int du_1du_2 p_1p_3e^{-yp_1(1 - 2u_1) - yp_3(1 - 2u_2) + \hbar p_1p_3( 1- 2u_2 - 2u_1) + \hbar p_1p_2 (2u_1) +  \hbar p_2p_3(-2u_2) }
			\end{aligned}
		\end{equation*}
		where $0<u_1<u_2<\frac{1}{2}$.
		Note that the two vertices $\bullet((\bullet)\bullet)$ and $(\bullet(\bullet))\bullet$ is symmetric under the change $u_1\leftrightarrow u_2$, therefore they can be combined to give a vertex
		\begin{equation*}
			\begin{aligned}
				& \bullet(\bullet(\bullet)) + (\bullet(\bullet))\bullet \\
				= &-4\hbar^2 c_1\int du_1du_2p_1p_2e^{\hbar\sum_{j<l}\bar{p}_j\bar{p}_l}e^{\hbar p_0p_1(1 - 2u_1) +\hbar p_0p_3(1 - 2u_3) + \hbar p_1p_3( 1- 2u_3 - 2u_1) + \hbar p_1p_2 (2u_1) + \hbar p_2p_3(-2u_3) } + c.c
			\end{aligned}
		\end{equation*}
		where $0<u_1,u_3<\frac{1}{2}$. We make the change of variable $u_3 \to 1 - u_3$, and define $u_0 = 0,u_2 = \frac{1}{2}$, and find that 
		
		\begin{equation*}
			m_{3,2} := (\bullet(\bullet))\bullet+ \bullet((\bullet)\bullet) =4\hbar^2 c_1\int_{\Delta^2} du_1du_3 p_1p_3e^{\hbar\sum_{j<l}2\bar{p}_j\bar{p}_l\bar{P}(u_l - u_j)+ \hbar\sum_{j<l}2p_jp_l P(u_l - u_j)} |_{p_3 \to -p_3}+ c.c
		\end{equation*}
		where the integration is taken over $\Delta^{2} = \{0  = u_0< u_1 < u_2 = \frac{1}{2} < u_3 < 1\}$.
		
		\paragraph{4. $((\bullet)\bullet)\bullet$}
		\begin{equation*}
			\begin{aligned}
				&p[h(h(d^2ze^{z(-\frac{1}{\hbar}y+p_1)})\star  e^{-yp_2})\star e^{-yp_3}]\\
				=& \hbar^2\int du_1du_2 p_2p_3e^{-yp_2(1 - 2u_2) - yp_3(1 - 2u_3) + \hbar p_2p_3(1 -2 u_2 + 2u_3) + \hbar p_3p_1 (2u_3) + \hbar p_2p_1(2u_2) }
			\end{aligned}
		\end{equation*}
		where $0<u_3<u_2<\frac{1}{2}$. We make the change of variable $u_2 \to 1- u_2, u_3 \to 1 - u_3$, define $u_0 = 0, u_1 = \frac{1}{2}$, and find that
		\begin{equation*}
			m_{3,3} := ((\bullet)\bullet)\bullet = 4\hbar^2 c_1 \int_{\Delta^3} du_2du_3 p_1p_2e^{\hbar\sum_{j<l}2\bar{p}_j\bar{p}_l\bar{P}(u_l - u_j)+ \hbar \sum_{j<l}2p_jp_l P(u_l - u_j)}|_{p_2,p_3\to -p_2,-p_3}+ c.c
		\end{equation*}
		where the integration is taken over $\Delta^3 = \{u_0 = 0, u_1  = \frac{1}{2}< u_2 < u_3 <1\}$.
		
		The above form of the vertices suggests to consider $P,\bar{P}$ as propagator of some physical model so that we can rewrite the vertices as the corresponding correlation function. 
		
		\subsection{A TQM inside Vasiliev equation}
		The symplectic group $Sp(n,\mathbb{R})$ acts on the symplectic space $(V,\omega)$ by linear transformation preserving $\omega$. For an element $g \in Sp(n,\mathbb{R})$, we consider field with twisted boundary condition
		\begin{equation*}
			\varphi(t+1) = g\varphi(t).
		\end{equation*}
		
		Specifically, we consider the case $\dim V = 4$ as in higher spin theory, and denote the dual basis by $y^{1},y^2,\bar{y}^{\dot{1}},\bar{y}^{\dot{2}}$. Recall that different from previous subsections \ref{TQM&Hoch}, the symplectic form we use in defining the star product is $\omega^{\alpha\beta} = 2\epsilon^{\alpha\beta},\omega^{\dot{\alpha}\dot{\beta}} = 2\epsilon^{\dot{\alpha}\dot{\beta}}$. We consider first the "holomorphic" part of the vertex $m_3$, which correspond to  a $\mathbb{Z}_2$ action defined by
		\begin{equation*}
			\bar{y}^{\dot{1}} \to -\bar{y}^{\dot{1}},\;\bar{y}^{\dot{2}} \to -\bar{y}^{\dot{2}},\; y^{1} \to y^1,\;y^2 \to y^2
		\end{equation*}
		In other word, the $\bar{y}$ directions have anti periodic condition and the $y$ direction is unchanged. Accordingly, the propagator will be different for the anti periodic boundary condition. The heat kernal in this case is
		\begin{equation*}
			\begin{aligned}
				K_t^{-}(\theta_1,\theta_2) & =  \frac{1}{\sqrt{4\pi t}} \sum_{n \in \mathbb{Z}}(-1)^{n}e ^{-\frac{(\theta_1 - \theta_2 + n)^2}{4t}}
				=\sum_{n\in\mathbb{Z}} e^{-\pi^2 (2n+1)^2 t}e^{ i (2n+1)\pi (\theta_1 - \theta_2)}.
			\end{aligned}
		\end{equation*}
		We find the propagator
		\begin{equation*}
			\begin{aligned}
				\bar{P}(\theta_1,\theta_2) & = \int_{0}^\infty \frac{\partial}{\partial \theta_1} K_t^-(\theta_1,\theta_2) dt
				 = \sum_{n\in \mathbb{Z}} \frac{i}{(2n+1)\pi } e^{i(2n+1)\pi (\theta_1 - \theta_2)}
			\end{aligned}
		\end{equation*}
		and we have
		\begin{equation*}
			\bar{P}(\theta_1,\theta_2) = \frac{1}{2}\quad \text{  for  }0< \theta_1 - \theta_2 < 1.
		\end{equation*}
		This is exactly the $\bar{P}$ we defined in last section. We find that the vertices found in last section can be calculated through the following correlation function
		\begin{equation*}
			\begin{aligned}
				m_{3,1} & = 4\hbar^2 c_1\left< \int_{\Delta^1}f^{(0)}_0(u_0)f^{(1)}_1(u_1)f^{(1)}_2(u_2)f^{(0)}_3(u_3) \right>\\
				m_{3,2} & = -4\hbar^2 c_1\left< \int_{\Delta^2}f^{(0)}_0(u_0)f^{(1)}_1(u_1)f^{(0)}_2(u_2)\pi(f_3)^{(1)}(u_3) \right>\\
				m_{3,3} & = 4\hbar^2 c_1\left< \int_{\Delta^3}f^{(0)}_0(u_0)f^{(0)}_1(u_1)\pi(f_2)^{(1)}(u_2)\pi(f_3)^{(1)}(u_3) \right>\\
			\end{aligned}
		\end{equation*}
		These vertices can be drawn as

		
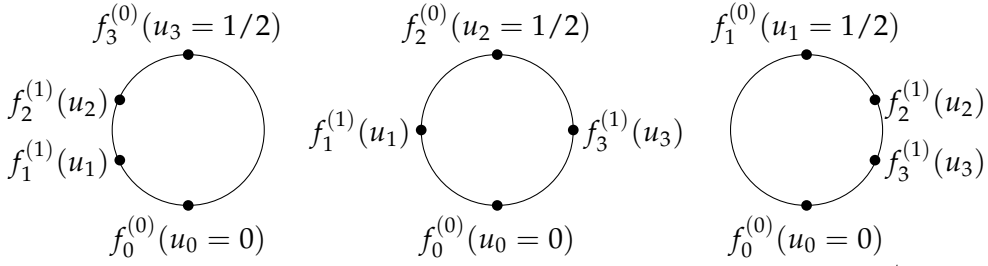
\begin{tikzpicture}
		\draw(0,0) circle (1);
		\fill (0,-1) circle (2pt) node[below] {$f^{(0)}_0(u_0 = 0)$};
		\fill (0,1) circle (2pt) node[above] {$f^{(0)}_3(u_3 = 1/2)$};
		\fill (-0.9,-0.4) circle (2pt) node[left] {$f^{(1)}_1(u_1)$};
		\fill (-0.9,0.4) circle (2pt) node[left] {$f^{(1)}_2(u_2)$};
		\end{tikzpicture}
		\begin{tikzpicture}
		\draw(0,0) circle (1);
		\fill (0,-1) circle (2pt) node[below] {$f^{(0)}_0(u_0 = 0)$};
		\fill (0,1) circle (2pt) node[above] {$f^{(0)}_2(u_2 = 1/2)$};
		\fill (-1,0) circle (2pt) node[left] {$f^{(1)}_1(u_1)$};
		\fill (1,0) circle (2pt) node[right] {$f^{(1)}_3(u_3)$};
		\end{tikzpicture}
		\begin{tikzpicture}
		\draw(0,0) circle (1);
		\fill (0,-1) circle (2pt) node[below] {$f^{(0)}_0(u_0 = 0)$};
		\fill (0,1) circle (2pt) node[above] {$f^{(0)}_1(u_1 = 1/2)$};
		\fill (0.9,-0.4) circle (2pt) node[right] {$f^{(1)}_3(u_3)$};
		\fill (0.9,0.4) circle (2pt) node[right] {$f^{(1)}_2(u_2)$};
		\end{tikzpicture}
		
	\noindent The appearance of $\pi$ above seems unnatural. We can define an operator $\frac{\overleftarrow{\partial}}{\partial \epsilon}$ by
		\begin{equation*}
			1\frac{\overleftarrow{\partial}}{\partial \epsilon} =0,\;\;\epsilon \frac{\overleftarrow{\partial}}{\partial \epsilon} = 1
		\end{equation*}
		Then we have the following formula
		\begin{equation*}
			\begin{aligned}
				\text{tr}(f_0\star m_{3,1}(f_1,f_2,f_3\epsilon)) &= 4\hbar^2 c_1\left< \int_{\Delta^1}f^{(0)}_0(u_0)f^{(1)}_1(u_1)f^{(1)}_2(u_2)f^{(0)}_3(u_3)\epsilon \frac{\overleftarrow{\partial}}{\partial \epsilon}\right>\\
				\text{tr}(f_0\star m_{3,2}(f_1,f_2\epsilon,f_3)) &= -4\hbar^2 c_1\left< \int_{\Delta^2}f^{(0)}_0(u_0)f^{(1)}_1(u_1)f^{(0)}_2(u_2)\epsilon f^{(1)}_3(u_3)\frac{\overleftarrow{\partial}}{\partial \epsilon}\right>\\
				\text{tr}(f_0\star m_{3,3}(f_1\epsilon,f_2,f_3)) &= 4\hbar^2 c_1\left< \int_{\Delta^3}f^{(0)}_0(u_0)f^{(0)}_1(u_1)\epsilon f^{(1)}_2(u_2) f^{(1)}_3(u_3)\frac{\overleftarrow{\partial}}{\partial \epsilon}\right>
			\end{aligned}
		\end{equation*}
		The construction for the anti-holomorphic part is similar, the TQM model is specified by the following $\mathbb{Z}_2$ action
	    \begin{equation*}
		 y^{1} \to -y^1,\;y^2 \to -y^2,\; \bar{y}^{\dot{1}} \to \bar{y}^{\dot{1}},\;\bar{y}^{\dot{2}} \to \bar{y}^{\dot{2}}
		\end{equation*}

		\section{Discussion}
		
		Our studies revel various structural aspects of higher spin theory. We give formulas to compute all order vertices in principle, which can be performed in ways by calculating tree level Feynman diagrams. This is obtained by applying homological perturbation theory to the choice of SDR data \eqref{hs_Vasiliev}. Note that the choice of this homotopy equivalence data is not unique. Although different choices will lead to equivalent $L_\infty$ algebra in principle, they might be related to nonlocal field redefinition, hence is of physical significance. A new class of shifted homotopy operators is introduced in \cite{Didenko:2018fgx,Gelfond:2018vmi}, and it is shown that proper choice can be used to decrease the level of non-locality of HS equations.\footnote{We thank M.~A.~Vasiliev for pointing out this important issue}. It would be important to explore to use of homotopy algebra techniques to fully analyze the locality of HS equations in all orders in interactions that is speculated in \cite{Gelfond:2018vmi}. 
		
		Here we would like to comment on the issue of AdS/CFT. Recall that the $AdS_4$ background is packed into a single $\mathfrak{hs}$ valued connection
		\begin{equation*}
			\Omega = \frac{1}{2}\varpi^{\alpha\beta}L_{\alpha\beta} + h^{\alpha\dot{\alpha}}P_{\alpha\dot{\alpha}} + \frac{1}{2}\bar{\varpi}^{\dot{\alpha}\dot{\beta}}\bar{L}_{\dot{\alpha}\dot{\beta}}.
		\end{equation*}
		It satisfies the equation 
		$$
		d_x\Omega  + \frac{1}{2}[\Omega,\Omega]_{\star} = 0.
		$$ 
		Moreover, $\Psi = \Omega$ is a solution to the full nonlinear HS equation, or equivalently, a Maurer-Cartan element. Therefore we can consider twisting the $L_\infty$ structure by $\Omega$
		\begin{equation*}
			\begin{aligned}
				l_1^{\Omega}(X) &= d_xX + \sum_{k\geq 0}\frac{1}{k!}l_{k+1}(\Omega^{\otimes k},X)\\
				l_{i}^{\Omega}(X_1,\dots,X_i) &= \sum_{k\geq 0}\frac{1}{k!}l_{k+i}(\Omega^{\otimes k},X_1,\dots,X_i),\; \text{ for }i>1
			\end{aligned}
		\end{equation*}
		We then have the Maurer-Cartan equation associated with this twisted $L_\infty$ structure
		\begin{equation*}
			\sum_{k\geq 1}\frac{1}{k!}l_k^{\Omega}(\Psi^{\otimes k}) = 0.
		\end{equation*}
		The physical meanings of this twisted Maurer-Cartan equation is that this is the original nonlinear HS equation fully expanded around the $AdS_4$ background $\Omega$. Specifically, the linear part
		\begin{equation*}
		l_1^{\Omega}(\Psi) = 0 
		\end{equation*}
		is the linearized higher spin equation \eqref{linearized_HS}. The holographic calculation of $n$-point function can be formally summarized as solving the above Maurer-Cartan equation with prescribed boundary behavior. Here we briefly describe it. We expand the field as
		\begin{equation*}
		\Psi = \Psi^{(1)} + \Psi^{(2)} + \Psi^{(3)} + \dots
		\end{equation*}
		They satisfy
		\begin{equation*}
		\begin{aligned}
		&l_1^{\Omega}(\Psi^{(1)}) = 0 \\
		&l_1^{\Omega}(\Psi^{(2)}) + \frac{1}{2!}l_2^{\Omega}(\Psi^{(1)},\Psi^{(1)}) = 0\\
		&l_1^{\Omega}(\Psi^{(3)}) + \frac{1}{3!}l_3^{\Omega}(\Psi^{(1)},\Psi^{(1)},\Psi^{(1)}) +  l_2^{\Omega}(\Psi^{(2)},\Psi^{(1)}) =0\\
		&\dots
		\end{aligned}
		\end{equation*}
		We can start with $\Psi^{(1)} = K\Psi_{\partial}$ and solve these equation order by order, where $K$ is the boundary to bulk propagator and $\Psi_{\partial}$ is the boundary source. Then the $n$-point function can be calculated (up to some normalization coefficient) by 
		\begin{equation*}
		\langle J_s J_{s_1}\dots J_{s_{n-1}}\rangle \sim \text{spin }s \text{ part } \frac{\delta^n\Psi^{(n-1)} }{\delta \Psi_{\partial}^{s_1} \delta \Psi_{\partial}^{s_2} \dots \delta \Psi_{\partial}^{s_{n-1}}}
		\end{equation*}
		where $\Psi_{\partial}^{s}$ refer to the spin-$s$ part of the boundary source $\Psi_{\partial}$. However, calculation in this way is cumbersome in practice. Simplification technique is developed to calculate the three point function in \cite{Giombi:2010vg}, where they essentially used $L_\infty$ morphism to solve the MC equation.  It would be very interesting to understand AdS/CFT for higher spin theory in terms of the algebraic structures we formulated in this paper,  in particular from the perspective of Koszul duality \cite{Costello:2017fbo,Costello:2016mgj}. Some perspectives of holography duality in the unfolded formulation related to homotopy structure is also discussed in \cite{Vasiliev:2012vf}.
		
			We also find that a topological quantum mechanics model hidden inside Vasiliev equation. This is not so surprising because the higher spin theory is based on the Weyl algebra which has quantum mechanical origin. This also implies that the Vasiliev theory itself is of first quantized nature. This closely resemble the case of string field theory, and further suggest its intimate relation with string theory.

		\Addresses

\begin{thebibliography}{99}
			
	\bibitem{Axelrod:1991vq} 
	  S.~Axelrod and I.~M.~Singer,
	  ``Chern-Simons perturbation theory,''
	  hep-th/9110056.
			
		
			\bibitem{Bekaert:2010hw}
			X.~Bekaert, N.~Boulanger and P.~Sundell, ``How higher-spin gravity surpasses the spin two barrier: no-go theorems versus yes-go examples,''
			Rev.\ Mod.\ Phys.\  {\bf 84} (2012) 987
			[arXiv:1007.0435 [hep-th]].
			
			\bibitem{Bershadsky:1993cx} 
	  M.~Bershadsky, S.~Cecotti, H.~Ooguri and C.~Vafa,
	  ``Kodaira-Spencer theory of gravity and exact results for quantum string amplitudes,''
	  Commun.\ Math.\ Phys.\  {\bf 165}, 311 (1994)
	  [hep-th/9309140].
	  
	  	\bibitem{Bonelli:2003kh}
			G.~Bonelli, ``On the tensionless limit of bosonic strings, infinite symmetries and higher spins,''
			Nucl.\ Phys.\ B {\bf 669} (2003) 159
			[hep-th/0305155].
			
			
			\bibitem{Crainic} M.~Crainic. "On the perturbation lemma, and deformations." arXiv preprint math/0403266 (2004).
			
			\bibitem{Costello:2017fbo} 
	  K.~Costello,
	  ``Holography and Koszul duality: the example of the $M2$ brane,''
	  arXiv:1705.02500 [hep-th].
	
	\bibitem{Costello:2012cy} 
	  K.~J.~Costello and S.~Li,
	  ``Quantum BCOV theory on Calabi-Yau manifolds and the higher genus B-model,''
	  arXiv:1201.4501 [math.QA].
	
			
			\bibitem{Costello:2016mgj} 
	  K.~Costello and S.~Li,
	  ``Twisted supergravity and its quantization,''
	  arXiv:1606.00365 [hep-th].
	
			
			\bibitem{Didenko:2014dwa} V.~E.~Didenko and E.~D.~Skvortsov, ``Elements of Vasiliev theory,'' [arXiv:1401.2975 [hep-th]].
			
			\bibitem{Didenko:2018fgx}
			V.~E.~Didenko, O.~A.~Gelfond, A.~V.~Korybut and M.~A.~Vasiliev, ``Homotopy Properties and Lower-Order Vertices in Higher-Spin Equations,'' arXiv:1807.00001 [hep-th].
			
			\bibitem{Feigin}B.~Feigin, G.~Felder, and B.~Shoikhet. "Hochschild cohomology of the Weyl algebra and traces in deformation quantization." Duke Mathematical Journal 127.3 (2005): 487-517.
			
			\bibitem{Getzler}E.~Getzler. "Lie theory for nilpotent $L_\infty$ algebras." Annals of mathematics (2009): 271-301.
			
			\bibitem{Getzler:1990}E.~Getzler, D.~S.~Jones. "$A_\infty$-algebras and the cyclic bar complex." Illinois J. Math 34.2 (1990): 256-283.
			
			\bibitem{Giombi:2009wh} S.~Giombi and X.~Yin, ``Higher Spin Gauge Theory and Holography: The Three-Point Functions,'' JHEP {\bf 1009}, 115 (2010) 
			[arXiv:0912.3462 [hep-th]].
			
			\bibitem{Giombi:2010vg}
			S.~Giombi and X.~Yin, ``Higher Spins in AdS and Twistorial Holography,''
			JHEP {\bf 1104} (2011) 086
			[arXiv:1004.3736 [hep-th]].
			
			\bibitem{Giombi:2012ms}
			S.~Giombi and X.~Yin,``The Higher Spin/Vector Model Duality,''
			J.\ Phys.\ A {\bf 46} (2013) 214003
			[arXiv:1208.4036 [hep-th]].
			
			\bibitem{Gelfond:2018vmi}
			O.~A.~Gelfond and M.~A.~Vasiliev, ``Homotopy Operators and Locality Theorems in Higher-Spin Equations,'' arXiv:1805.11941 [hep-th].
			
			
			\bibitem{Huebschmann} J.~Huebschmann. On the construction of $A_\infty$ structures. Georgian Mathematical Journal, 17(1), pp. 161-202 (2010). 
			
			\bibitem{Huebschmann:2002}J.~Huebschmann and J.~Stasheff. "Formal solution of the master equation via HPT and deformation theory." Forum mathematicum. Vol. 14. No. 6. Berlin; New York: De Gruyter, c1989-, 2002.
			
			\bibitem{Kontsevich:1997vb}
			M.~Kontsevich, ``Deformation quantization of Poisson manifolds. 1.,''
			Lett.\ Math.\ Phys.\  {\bf 66} (2003) 157
			[q-alg/9709040].
			
			\bibitem{Keller}B.~Keller. Introduction to $A$-infinity algebras and modules. 
			Homology Homotopy Appl. 3 (2001), no. 1, 1--35.
			
			
			\bibitem{Klebanov:2002ja}
			I.~R.~Klebanov and A.~M.~Polyakov, ``AdS dual of the critical O(N) vector model,''
			Phys.\ Lett.\ B {\bf 550} (2002) 213
			[hep-th/0210114].
			
			\bibitem{Kajiura:2003ax}
			H.~Kajiura, ``Noncommutative homotopy algebras associated with open strings,''
			Rev.\ Math.\ Phys.\  {\bf 19} (2007) 1
			[math/0306332 [math-qa]].
			
			
			\bibitem{Kajiura:2004xu}
			H.~Kajiura and J.~Stasheff, ``Homotopy algebras inspired by classical open-closed string field theory,''
			Commun.\ Math.\ Phys.\  {\bf 263} (2006) 553
			[math/0410291 [math-qa]].
			
			\bibitem{Lada:1992wc}
			T.~Lada and J.~Stasheff, ``Introduction to SH Lie algebras for physicists,''
			Int.\ J.\ Theor.\ Phys.\  {\bf 32} (1993) 1087
			[hep-th/9209099].
			
	\bibitem{Si-index} 
	  R.~E.~Grady, Q.~Li and S.~Li,
	  ``Batalin–Vilkovisky quantization and the algebraic index,''
	  Adv.\ Math.\  {\bf 317}, 575 (2017)
	  [arXiv:1507.01812 [math.QA]].
			
			\bibitem{Markl}M.~Markl. Transferring $A_\infty$ (strongly homotopy associative) structures. Rend. Circ. Mat. Palermo (2) Suppl., (79):139–151, 2006. 139151.
			
			\bibitem{Shoikhet:2000gw}
			B.~Shoikhet, A proof of the Tsygan formality conjecture for chains, Adv. Math. 179 (2003), 7 – 37. 
			
			\bibitem{Sezgin:2000hr}
			E.~Sezgin and P.~Sundell,``On curvature expansion of higher spin gauge theory,''
			Class.\ Quant.\ Grav.\  {\bf 18} (2001) 3241
			[hep-th/0012168].
			
			\bibitem{Sezgin:2002ru} E.~Sezgin and P.~Sundell, ``Analysis of higher spin field equations in four-dimensions,'' JHEP {\bf 0207}, 055 (2002) 
			[hep-th/0205132].
			
			
			
			\bibitem{Sagnotti:2003qa}
			A.~Sagnotti and M.~Tsulaia, ``On higher spins and the tensionless limit of string theory,''
			Nucl.\ Phys.\ B {\bf 682} (2004) 83
			[hep-th/0311257].
			
			\bibitem{Sharapov:2017yde} A.~A.~Sharapov and E.~D.~Skvortsov,
			``Formal higher-spin theories and Kontsevich-Shoikhet-Tsygan formality,'' Nucl.\ Phys.\ B {\bf 921}, 538 (2017) 
			[arXiv:1702.08218 [hep-th]].
			
			
			\bibitem{Vasiliev:1988sa}
			M.~A.~Vasiliev,``Consistent Equations for Interacting Massless Fields of All Spins in the First Order in Curvatures,''
			Annals Phys.\  {\bf 190} (1989) 59.
			
			\bibitem{Vasiliev:1990en}
			M.~A.~Vasiliev, ``Consistent equation for interacting gauge fields of all spins in (3+1)-dimensions,''
			Phys.\ Lett.\ B {\bf 243} (1990) 378.
			
			\bibitem{Vasiliev:1990vu}
			M.~A.~Vasiliev,``Properties of equations of motion of interacting gauge fields of all spins in (3+1)-dimensions,''
			Class.\ Quant.\ Grav.\  {\bf 8} (1991) 1387.
			
			\bibitem{Vasiliev:1992av}
			M.~A.~Vasiliev, ``More on equations of motion for interacting massless fields of all spins in (3+1)-dimensions,''
			Phys.\ Lett.\ B {\bf 285} (1992) 225.
			
			\bibitem{Vasiliev:1999ba} M.~A.~Vasiliev, ``Higher spin gauge theories: Star product and $AdS$ space,''  In *Shifman, M.A. (ed.): The many faces of the superworld* 533-610 
			
			

			
						\bibitem{Vasiliev:2005zu}
			M.~A.~Vasiliev,
			``Actions, charges and off-shell fields in the unfolded dynamics approach,''
			Int.\ J.\ Geom.\ Meth.\ Mod.\ Phys.\  {\bf 3} (2006) 37
			[hep-th/0504090].
			

						\bibitem{Vasiliev:2012vf}
			M.~A.~Vasiliev, ``Holography, Unfolding and Higher-Spin Theory,''
			J.\ Phys.\ A {\bf 46} (2013) 214013
			[arXiv:1203.5554 [hep-th]].
			
			\bibitem{Witten:1992fb} 
			
		

  E.~Witten,
  ``Chern-Simons gauge theory as a string theory,''
  Prog.\ Math.\  {\bf 133}, 637 (1995)
  [hep-th/9207094].
			
			\bibitem{Zwiebach:1992ie}
			B.~Zwiebach, ``Closed string field theory: Quantum action and the B-V master equation,''
			Nucl.\ Phys.\ B {\bf 390} (1993) 33
			[hep-th/9206084].
			
			\bibitem{Zwiebach:1997fe} 
	  B.~Zwiebach,
	  ``Oriented open - closed string theory revisited,''
	  Annals Phys.\  {\bf 267}, 193 (1998)
	  [hep-th/9705241].
			
		\end{thebibliography}
	\end{document}